\documentclass{iopjournal}

\usepackage{amsmath}
\usepackage{amssymb}
\usepackage{float}

\usepackage{tikz}
\usetikzlibrary{positioning,arrows.meta}

\begin{document}

\pagestyle{fancy}
\fancyhf{}

\fancyhead[L]{Charles Wood}
\fancyhead[R]{Information Theory: An X-ray Microscopy Perspective}

\fancyfoot[C]{\thepage}

\articletype{Paper} 

\title{Information Theory: An X-ray Microscopy Perspective}

\author{Charles Wood$^1$\orcid{0009-0007-4614-2695}}

\affil{$^1$Future Technology Centre, School of Electrical and Mechanical Engineering, University of Portsmouth, UK}

\email{charles.wood@port.ac.uk}


\begin{abstract}

\medskip
X-ray microscopy (XRM) is routinely used to obtain three-dimensional information on internal
microstructure, but the imaging pipeline introduces noise, redundancy and information loss at
multiple stages. This paper adopts an explicitly information-theoretic standpoint and treats the
XRM workflow as a sequence of processing steps acting on a finite information budget. Using entropy, mutual information and Kullback--Leibler divergence, we quantify how acquisition, denoising, alignment, sparse-angle sampling, dose variation and reconstruction each reshape the statistical structure of projection data and reconstructed volumes. Case studies based on the publicly available \emph{Walnut~1} dataset illustrate how different denoising algorithms redistribute grey-level occupancy, how misalignment and registration affect mutual information across projections, how sparse-angle sampling imposes an information bottleneck, how dose constrains the fraction of entropy attributable to meaningful structure, and how filtered backprojection and iterative reconstruction act as distinct information-shaping operators. Building on these results, we summarise the XRM workflow using a unified information budget, examine an empirical dose--information relation for Poisson--Gaussian projection data, and show that mutual information provides a reconstruction-agnostic indicator of fidelity. An empirical information--degradation hierarchy is proposed to rank the informational impact of pipeline components. Overall, the analysis supports viewing XRM as an information-processing system and provides a quantitative framework for comparing and optimising imaging protocols, particularly in low-dose or time-constrained settings. All expressions introduced here are leading-order, empirically calibrated formulations intended to be self-contained for protocol comparison within the stated estimator conventions; no result in the present paper depends on external theoretical developments for its interpretation.

\end{abstract}

\section{Introduction}

\medskip

As with most things, beginning with ``why'', then ``how'', and finally ``what'' provides a useful structure. X-ray microscopy (XRM) is a powerful, non-destructive imaging technique used to investigate the internal microstructure of objects at high spatial resolution. It plays a critical role across disciplines such as materials science, biology, and medicine by enabling detailed three-dimensional (3D) representations of samples~\cite{Kak2001,Stock2008}. However, the XRM workflow, encompassing data acquisition, image reconstruction, and post-processing, introduces various sources of noise, redundancy, and information loss~\cite{Boas2012,Hsieh2015}. These distortions degrade image quality and, ultimately, limit the scientific insight that can be extracted from the data.

\medskip
\noindent Information theory, developed by Claude Shannon in the mid-20th century, provides a mathematical framework for quantifying information content and uncertainty in data~\cite{Shannon1948}. Originally applied to telecommunications and data compression, it is increasingly relevant in imaging sciences~\cite{Cover2006}. Key concepts such as entropy, which measures uncertainty or randomness in a dataset, and mutual information, which quantifies the information shared between two signals, offer rigorous, quantitative tools for evaluating image quality, noise levels, and reconstruction fidelity~\cite{Gonzalez2018}. In the context of XRM, entropy can be used to characterise noise in projection images and reconstructions, or to evaluate the impact of denoising algorithms. Lower entropy may indicate higher image fidelity or reduced randomness, whereas higher entropy may reflect either noise or increased structural complexity. Mutual information between input and reconstructed data offers a robust metric for assessing how well essential features are preserved during reconstruction, particularly when evaluating different algorithms, noise models, or acquisition settings~\cite{Barrett2004}.

\medskip
\noindent This paper proposes an information-theoretic framework for analysing the XRM pipeline. By systematically applying entropy and information-based metrics to each stage: acquisition, reconstruction, and post-processing, we aim to quantify how information flows through the system, to identify where and how information is lost, distorted, or preserved, and to explore the trade-offs between image quality, radiation dose, and computational complexity. In doing so, we seek to establish new quantitative metrics for evaluating and optimising imaging protocols. This approach is further extended to incorporate noise characterisation, theoretical limits on information throughput, and case studies using both simulated and experimental datasets.

\medskip
\noindent The present paper develops and validates an applied information-theoretic view of XRM using standard metrics (entropy, mutual information and Kullback--Leibler divergence) to analyse a representative micro-CT workflow. The focus is on how these quantities behave in practice across acquisition, alignment, sparse-angle sampling, dose variation and reconstruction, using case studies on the \emph{Walnut~1} dataset to construct an empirical information budget for XRM. A separate theoretical treatment develops an operator-level formalism for imaging systems; however, no result in the present paper requires that formalism for interpretation, and all relations used here are stated only as empirically calibrated, leading-order proxies within the estimator conventions declared in Section~7.

\medskip
\noindent Throughout this paper, terms such as \emph{information}, \emph{entropy}, and \emph{information budget} are used strictly to denote measured statistical variability under fixed discretisation, masking, and normalisation conventions. They do not imply conserved quantities, task-independent meaning, physical invariants, or estimator-free notions of information. All numerical values, trends, and orderings reported are therefore conditional on these conventions and are intended solely for like-for-like comparison of imaging protocols analysed under identical preprocessing and estimation choices.

\section{Information Metrics for X-ray Microscopy}

\medskip
Information theory provides a robust framework for analysing and optimising the XRM process. This section introduces three key metrics as they are used in the present work: entropy, mutual information and Kullback--Leibler divergence, with channel capacity included as a conceptual ceiling. These quantities provide the working basis for quantifying information throughout the imaging workflow and for evaluating how much information is captured, lost or transformed during acquisition, reconstruction and post-processing.

\subsection{Entropy and Noise in Projection Data}

\medskip
Entropy is a measure of uncertainty or randomness within a system. In imaging, it quantifies the information content of image data and provides insight into the degree of noise or disorder present at various stages of the workflow~\cite{Shannon1948, Cover2006, Gonzalez2018}.

\medskip
\noindent For a discrete random variable $X$ with probability distribution $p(x)$, the Shannon entropy $H(X)$ is defined as
\begin{equation}
  H(X) = - \sum_{x} p(x)\,\log_2 p(x).
  \label{eq:entropy}
\end{equation}

\medskip
\noindent\textbf{Entropy definition used in this paper.}
Unless stated otherwise, all entropies reported in Sections~7.1--7.7 are \emph{discrete} Shannon entropies computed from normalised intensity histograms,
\[
H_{\Delta}(X) \;=\; -\sum_{b=1}^{B} p_b \log_2 p_b,
\]
where $p_b$ is the empirical probability mass in histogram bin $b$, $B$ is the fixed bin count, and $\Delta$ denotes the induced discretisation (bin width and intensity normalisation).
Accordingly, numerical entropy values depend on the chosen normalisation, masking, and binning; throughout the case studies we keep these choices fixed within each experiment to ensure like-for-like comparisons.
When we discuss \emph{differential} entropy approximations (e.g.\ Section~6.2), this is explicitly labelled and used only to describe leading-order \emph{scaling} with dose, not to predict the absolute discrete entropies reported in the tables.

\medskip
\noindent\textbf{Discrete vs.\ differential entropy.}
The histogram entropy $H_{\Delta}$ defined above is the primary quantity reported in this paper.
When we later write differential-entropy expressions (e.g.\ Section~6.2), these are used only to describe
\emph{qualitative scaling} with dose under an approximately Gaussian fluctuation model.
No step in the paper identifies $H_{\Delta}$ with a differential entropy $h$, and derivatives such as
$\mathrm{d}H/\mathrm{d}\lambda$ should be read as derivatives of the \emph{differential-entropy proxy} used for scaling,
not as derivatives of the discrete histogram entropies reported in Tables~\ref{tab:entropy_kl}--\ref{tab:dose_metrics}.

\medskip
\noindent This expression gives the expected number of bits needed to encode the outcomes of $X$. In X-ray microscopy (XRM), the variable $X$ typically represents the intensity values of an image's pixels or voxels. The entropy of an image, therefore, reflects how pixel intensities are distributed and serves as a proxy for both information content and noise~\cite{Hsieh2015, Shannon1948}.

\medskip
\noindent Entropy can be used to analyse different stages of the XRM pipeline. During data acquisition, it provides a quantitative estimate of the noise present in the raw projection images~\cite{Barrett2004,Siewerdsen2001}. Higher entropy implies greater randomness and potentially lower signal fidelity. During reconstruction, entropy helps evaluate the extent to which structural information has been preserved or degraded by the algorithm~\cite{Barrett2004}. Reconstruction methods that effectively suppress noise while retaining signal will generally result in lower entropy compared to noisy or artefact-laden reconstructions~\cite{Barrett2004, Nuyts2013}. In post-processing steps such as denoising and contrast enhancement, changes in entropy can indicate the impact of those operations on image quality. Ideally, denoising should reduce entropy by removing stochastic variation while preserving meaningful features~\cite{Gonzalez2018}.

\medskip
\noindent The entropy of projection data is also shaped by several key system-level parameters. Detector noise contributes directly to entropy by introducing random fluctuations that obscure underlying structure~\cite{Foi2008,Ma2012}. Spatial resolution affects entropy through the level of structural detail captured. Higher resolution generally increases entropy by revealing fine-scale variations in the sample, though this effect is modulated by the signal-to-noise ratio~\cite{Barrett2004}. Bit depth, which defines the number of grey levels the detector can record, places an upper limit on entropy~\cite{Hsieh2015}. For example, an 8-bit detector cannot exceed 8~bits of entropy per pixel. Radiation dose indirectly influences entropy via its impact on photon statistics. Lower doses reduce the number of detected photons, increasing stochastic noise and thereby raising entropy. Conversely, higher doses improve signal quality but may be constrained by concerns around sample damage, safety, or throughput~\cite{Boas2012, Zhang2014}. Understanding how these parameters interact provides critical context for interpreting entropy measurements in both experimental and simulated datasets.

\medskip
\noindent Entropy will be used throughout this paper as a quantitative metric for assessing information quality in projection data and reconstructed volumes. In the following subsection, we turn to mutual information as a complementary tool for comparing datasets and evaluating reconstruction fidelity.

\subsection{Mutual Information and Reconstruction Fidelity}

\medskip
While entropy quantifies the uncertainty within a single dataset, mutual information provides a measure of the statistical dependence between two datasets~\cite{Cover2006}. In the context of X-ray microscopy, mutual information can be used to evaluate how much of the original information captured during projection acquisition is retained in the reconstructed image~\cite{Barrett2004}. This is particularly important for assessing the performance of reconstruction algorithms, registration accuracy, or the effect of post-processing steps on data integrity.

\medskip
\noindent Formally, the mutual information $I(X;Y)$ between two discrete random variables $X$ and $Y$, with joint probability distribution $p(x,y)$ and marginal distributions $p(x)$ and $p(y)$, is defined as~\cite{Cover2006, Kullback1951}
\begin{equation}
  I(X;Y) = \sum_{x,y} p(x,y)\,\log_2 \frac{p(x,y)}{p(x)\,p(y)}.
  \label{eq:mi}
\end{equation}
This quantity represents the reduction in uncertainty about one variable given knowledge of the other. If $X$ and $Y$ are independent, their mutual information is zero. If they are perfectly correlated, mutual information reaches its maximum and equals the entropy of either variable.

\medskip
\noindent In XRM applications, $X$ might represent the intensity values of the projection data and $Y$ the corresponding values in a reconstructed or post-processed image. Mutual information then quantifies the degree to which the reconstruction preserves the structural information present in the original measurement~\cite{Barrett2004}. This makes it a powerful tool for comparing different reconstruction algorithms, evaluating registration quality in multi-view or multimodal datasets, or assessing the impact of denoising and artefact removal techniques.

\noindent Unlike traditional image similarity metrics such as mean squared error (MSE) or structural similarity index (SSIM), mutual information does not require pixel-wise correspondence and is robust to intensity scaling or nonlinear transformations~\cite{Gonzalez2018}. This makes it particularly suitable for comparing datasets acquired under different exposure settings, energy levels, or contrast mechanisms. Mutual information has been widely used in medical image registration~\cite{Maes1997,Viola1997}, but its broader utility in XRM remains underexplored.

\medskip
\noindent In this paper, we will use mutual information to quantify the fidelity of reconstructed images relative to reference datasets, either from simulated ground truth or high-dose acquisitions. This allows us to assess whether increased entropy in a reconstruction corresponds to meaningful structural complexity or simply added noise, and to compare how different acquisition parameters and reconstruction pipelines influence information retention.

\subsection{Kullback--Leibler Divergence and Distributional Change}

\medskip
The Kullback--Leibler (KL) divergence is a fundamental concept in information theory that quantifies the difference between two probability distributions~\cite{Cover2006, Kullback1951}. While entropy and mutual information describe uncertainty and shared information, KL divergence provides a directional measure of how one distribution diverges from another. In the context of X-ray microscopy (XRM), KL divergence can be used to assess how closely reconstructed, denoised, or processed images match an expected or reference distribution~\cite{Viola1997,Sijbers2004,Howells2009}, such as a known ground truth or a simulated model.

\medskip
\noindent For two discrete probability distributions $P(x)$ and $Q(x)$ defined over the same domain $\mathcal{X}$, the KL divergence from $Q$ to $P$ is given by~\cite{Cover2006,Kullback1951}
\begin{equation}
  D_{\mathrm{KL}}(P\|Q) = \sum_{x \in \mathcal{X}} P(x)\,\log_2 \frac{P(x)}{Q(x)}.
  \label{eq:kl}
\end{equation}
This expression represents the additional number of bits required to encode samples from $P$ using a code optimised for $Q$, rather than one optimised for $P$ itself. The KL divergence is non-negative and equals zero only when $P$ and $Q$ are identical.

\medskip
\noindent In practical imaging scenarios, $P$ might represent the empirical distribution of intensity values in a high-fidelity reference image, while $Q$ represents the distribution from a lower-quality reconstruction or a version of the data after denoising. The KL divergence then quantifies how much structural or textural information has been lost or altered in the transformation. Since KL divergence is asymmetric, it can highlight directional biases introduced by certain algorithms, such as oversmoothing, artefact amplification, or contrast compression~\cite{Barrett2004}.

\medskip
\noindent KL divergence is especially useful when comparing probabilistic representations of image data, such as intensity histograms, noise models, or learned priors~\cite{Foi2008,Ma2012}. For example, it can be applied to assess whether a denoising algorithm alters the statistical distribution of the signal, or to evaluate how well a reconstruction algorithm preserves rare features that are present in the original data but smoothed out in post-processing~\cite{Sijbers2004}. In simulated workflows, it allows quantitative comparison between a known input distribution and the result of a simulated acquisition and reconstruction pipeline.

\medskip
\noindent While KL divergence is not a direct measure of visual similarity, it complements metrics like mutual information by offering a way to detect systematic shifts in data representation. In this paper, we apply KL divergence to evaluate distributional consistency between raw and processed data, particularly in assessing how denoising and reconstruction algorithms influence the statistical structure of the underlying signal.

\subsection{Channel Capacity and Information Limits in Imaging Systems}

\medskip
\noindent Channel capacity, a central concept in information theory, defines the maximum rate at which
information can be reliably transmitted through a communication channel \cite{Shannon1948, Cover2006}.
When applied to imaging systems such as X-ray microscopy (XRM), the concept of channel capacity can be
used to frame the theoretical limits on how much information the imaging system can extract from the
sample, given its physical and operational constraints.

\medskip
\noindent In Shannon's original formulation, the channel capacity $C$ of a discrete memoryless channel with
input $X$, output $Y$, and conditional distribution $p(y|x)$, is given by the maximum mutual
information between input and output \cite{Shannon1948, Cover2006}:
\begin{equation}
C = \max_{p(x)} I(X;Y).
\label{eq:capacity}
\end{equation}
This quantity represents the upper bound on the amount of information (in bits) that can be transmitted
per use of the channel without error, assuming optimal encoding.

\medskip
\noindent In the context of XRM, the imaging system can be viewed as an information channel in which the underlying structure of the sample is mapped to a finite set of noisy measurements. The effective channel capacity is constrained by photon statistics, detector quantisation, spatial sampling, and
system noise. These constraints are imposed at acquisition and cannot be relaxed by downstream processing.

\medskip
\noindent An important consequence is that reconstruction quality is bounded once the acquisition parameters are fixed. Improvements in reconstruction algorithms may redistribute error, suppress noise, or impose prior structure, but they cannot increase the amount of independent information captured by the
measurement process. In this sense, reconstruction operates within a capacity-limited regime defined upstream.

\medskip
\noindent This perspective clarifies several common pathologies in imaging workflows. Increasing reconstruction complexity beyond the acquisition-limited capacity yields diminishing returns and may produce apparent improvements in visual quality that do not correspond to increased information retention. Similarly, oversampling in space or bit depth without a corresponding increase in photon statistics does not
increase effective capacity and instead redistributes noise.

\medskip
\noindent In this paper, \emph{channel capacity} is used only as a conceptual ceiling: it emphasises that acquisition physics bounds the information that downstream computation can exploit. We do not claim that Eq.~\eqref{eq:capacity} applies directly to the XRM pipeline (which is typically continuous, correlated, and governed by a physical forward operator rather than a discrete memoryless channel). Accordingly, we use ``capacity-limited'' as shorthand for \emph{acquisition-limited}: once dose, sampling, and detector constraints are fixed, reconstruction can at best redistribute the measured information under additional modelling assumptions.

\section{Information Flow and Entropy Across the X-ray Microscopy Pipeline}

\medskip
This section traces the flow of information through the X-ray microscopy (XRM) pipeline, from initial data acquisition to final post-processing. At each stage, we analyse how entropy evolves and assess where information is gained, lost, or transformed. By mapping these dynamics, we aim to establish a quantitative framework that highlights bottlenecks, inefficiencies, and opportunities for optimisation. This analysis is intended to apply broadly across the field of X-ray microscopy, encompassing full-field and scanning modalities, laboratory- and synchrotron-based systems, and both micro- and nano-scale imaging techniques.

\subsection{Data Acquisition}

\medskip
The data acquisition stage is the first point at which entropy enters the system. At this point, the total entropy in the projection data reflects a combination of structural complexity in the sample and noise introduced during detection. This entropy forms the baseline against which all downstream information gain or loss will be measured.

\medskip
\noindent Entropy at this stage originates from two principal sources. The first is structural entropy, which represents the spatial variation in X-ray attenuation arising from the internal features of the sample. Samples with high heterogeneity or fine structure generate greater variation in detected intensities, leading to increased entropy~\cite{Barrett2004}. The second is noise entropy, which results from statistical fluctuations in photon detection (commonly modelled as Poisson noise) and from detector-related factors such as electronic readout noise and thermal noise~\cite{Siewerdsen2001,Foi2008}. These phenomena have been characterised in empirical CT and radiographic imaging studies. Reduced exposure (for example, lower mAs or dose) increases noise variance in projection data, consistent with Poisson-dominated behaviour at reduced photon counts~\cite{Siewerdsen2001,Zhang2014}. Comprehensive reviews have detailed how readout electronics and thermal processes degrade image fidelity, and models have been developed to quantify mixed Poisson--Gaussian noise in X-ray projection images~\cite{Foi2008,Ma2012}, enabling improved estimation of entropy contributions under low-dose conditions.

\medskip
\noindent The balance between structural and noise entropy is heavily influenced by acquisition parameters. Increasing the radiation dose improves the signal-to-noise ratio by increasing the number of detected photons, thereby reducing the relative contribution of noise entropy~\cite{Siewerdsen2001,Zhang2014}. However, practical limits such as scan time, sample damage, and system throughput often constrain the maximum dose that can be applied. Lower dose acquisitions result in noisier data, which increases total entropy but reduces its proportion of useful content~\cite{Boas2012,Zhang2014}.

\medskip
\noindent Other acquisition parameters also shape the entropy profile of projection data. Bit depth defines the number of discrete grey levels available for digitisation, which in turn places a theoretical upper bound on the entropy that can be represented in the image~\cite{Hsieh2015}. Spatial resolution influences the amount of structural detail that can be captured. While higher resolution allows for the detection of finer features and increases structural entropy, it also tends to amplify noise per voxel, especially under limited dose conditions~\cite{Barrett2004}. Additionally, acquisition settings such as filtration, exposure time, and beam energy modulate the X-ray spectrum and intensity distribution, altering the signal-to-noise ratio and the corresponding entropy landscape~\cite{Siewerdsen2001}.

\medskip
\noindent Entropy can be measured directly from the raw projection images by analysing the grey-level distribution using standard entropy estimators. These include histogram-based methods, differential entropy calculations, and kernel density approaches~\cite{Gonzalez2018}. Such measurements provide an empirical summary of the total uncertainty in the data. However, interpretation requires care. An increase in entropy could result from either more structural detail or more noise, and the two are not always easily separable. Repeating entropy measurements under different acquisition settings can help identify which factors contribute most to the observed entropy profile.

\medskip
\noindent Understanding the entropy introduced during acquisition is essential because it defines the upper limit of what can be recovered during reconstruction. Any meaningful signal not captured here cannot be retrieved later, and any noise introduced at this stage may propagate or amplify downstream. This makes the acquisition stage the single most critical contributor to the information content available in the final reconstruction~\cite{Barrett2004}. These results show that entropy is a sensitive and quantitative indicator of noise levels in projection data. When used alongside ground truth or high-fidelity references, entropy provides an objective means of evaluating the impact of acquisition settings and denoising techniques~\cite{Barrett2004,Siewerdsen2001}. This establishes entropy as a foundational metric for the more complex analyses to follow.

\subsection{Image Formation}

\medskip
After acquisition, the raw projections are formed as the output of X-ray interactions with the sample and the characteristics of the detector. At this stage, entropy is shaped by both the physical properties of the beam--sample interaction and the design of the imaging system. The resulting projections encode both useful structural information and various artefacts that influence the entropy profile.

\medskip
\noindent Beam hardening introduces nonlinearity into the intensity distribution, as lower-energy photons are more strongly absorbed than higher-energy ones. This effect leads to non-uniform contrast, particularly in thick or multi-material samples, and can increase entropy by distorting otherwise structured attenuation patterns~\cite{Boas2012,Hsieh2015}. Similarly, scatter contributes diffuse signals across the detector, blurring local features and introducing spatial correlations that raise noise entropy~\cite{Boas2012,Hsieh2015}.

\medskip
\noindent Sample composition also plays a critical role. High-contrast boundaries or interfaces between different materials enhance structural entropy, as they generate significant intensity variations in the projections. Conversely, homogeneous or weakly absorbing regions may result in flatter histograms and lower structural entropy. In such cases, noise contributes a larger proportion of the total entropy~\cite{Barrett2004}.

\medskip
\noindent Detector characteristics further modulate the entropy profile. The modulation transfer function (MTF) defines the system's ability to resolve detail, influencing how spatial features are recorded and how entropy is distributed across frequency space~\cite{Barrett2004}. A detector with poor resolution may suppress fine detail, reducing structural entropy but leaving noise entropy unchanged or even amplified~\cite{Siewerdsen2001}. Similarly, detector blur and charge diffusion can smooth out sharp features, lowering spatial entropy while preserving global histogram entropy~\cite{Hsieh2015}.

\medskip
\noindent Non-idealities such as detector drift, defective pixels, or non-uniform response can introduce additional structured artefacts into the projection data. These artefacts raise entropy in a way that may mimic structural features, complicating later reconstruction and interpretation~\cite{Boas2012,Sijbers2004}.

\medskip
\noindent Entropy analysis of projection images provides an empirical summary of how much uncertainty has been introduced or preserved during image formation. Comparing entropy across detector types, beam settings, or sample compositions can help identify setups that optimise structural information while minimising artefactual contributions~\cite{Barrett2004}.

\subsection{Image Reconstruction}

\medskip
\noindent Image reconstruction transforms the raw projection data into spatially resolved volumetric or planar representations of the sample. This stage is a critical juncture in the information pipeline, where entropy can either be preserved, redistributed, or lost entirely, depending on the reconstruction algorithm and its assumptions.

\medskip
\noindent Filtered backprojection (FBP) methods, widely used for their computational efficiency, apply convolution filters to projection data before backprojecting across all angles. While fast and deterministic, FBP assumes idealised sampling conditions and can amplify noise or artefacts when data is sparse or under-sampled~\cite{Kak2001,Hsieh2015}. These effects manifest as increased high-frequency components that may raise noise entropy or obscure structural details~\cite{Siewerdsen2001}.

\medskip
\noindent Iterative reconstruction methods, such as algebraic reconstruction techniques (ART) and model-based iterative reconstruction (MBIR), refine the reconstruction through successive approximations, incorporating prior knowledge or noise models. These methods often yield lower entropy volumes with sharper features, reflecting a higher signal-to-noise ratio and more effective suppression of artefacts~\cite{Barrett2004}. However, the reduction in entropy is not always indicative of preserved information; over-regularisation or inappropriate priors can eliminate subtle but meaningful features, replacing them with artificially smooth structures~\cite{Kak2001}.

\medskip
\noindent Information loss also arises from the interpolation steps used in grid alignment, voxel sampling, or angular rebinning. Each of these operations introduces smoothing or aliasing effects that suppress high-frequency components, reducing structural entropy~\cite{Hsieh2015}. The degree of loss depends on interpolation kernel, sampling density, and noise characteristics.

\medskip
\noindent Reconstruction artefacts, including streaks, rings, and truncation errors, also contribute structured noise that can increase entropy in misleading ways. While they may visually resemble signals, these artefacts introduce false complexity and reduce the mutual information between the reconstruction and the true object~\cite{Boas2012,Hsieh2015}.

\medskip
\noindent Thus, reconstruction methods must be evaluated not only in terms of visual quality, but also through their effect on entropy and information retention. Comparative entropy analysis across methods can reveal how much useful information has been recovered versus how much distortion or redundancy has been introduced~\cite{Barrett2004}.

\subsection{Post-Processing and Enhancement}

\medskip
\noindent Post-processing encompasses a range of techniques aimed at refining the reconstructed image, including denoising, artefact correction, contrast enhancement, and spatial alignment. Each of these operations modifies the entropy profile, either by discarding irrelevant uncertainty or, in some cases, unintentionally suppressing real signals.

\medskip
\noindent Denoising algorithms reduce noise entropy by removing stochastic fluctuations in intensity. Simple methods like Gaussian blurring decrease overall entropy but often sacrifice edge fidelity~\cite{Gonzalez2018}. More advanced filters, such as bilateral filtering or non-local means, attempt to preserve structural entropy while suppressing random noise~\cite{Gonzalez2018,Foi2008}. The effectiveness of a denoising method can be evaluated by measuring entropy before and after processing and computing the mutual information retained from the original image~\cite{Barrett2004}.

\medskip
\noindent Registration and alignment steps, particularly in multimodal or time-series imaging, aim to improve spatial correspondence between datasets. However, interpolation errors or transformation artefacts can introduce new forms of entropy~\cite{Gonzalez2018}. The trade-off between spatial coherence and interpolation loss should be assessed using entropy-based metrics~\cite{Barrett2004}.

\medskip
\noindent Artefact correction routines, such as ring removal or beam hardening compensation, target structured noise that inflates entropy without adding useful information. Their success can be quantified by observing reductions in non-informative entropy components, ideally without suppressing real features~\cite{Boas2012,Sijbers2004}.

\medskip
\noindent Overall, post-processing should aim to minimise noise and artefacts while preserving or enhancing mutual information between the reconstruction and the original structure. Entropy analysis provides a consistent framework to assess the quality and cost of each enhancement step, supporting decisions about which processing methods offer the greatest information return~\cite{Barrett2004}.

\section{Noise Characterisation and Its Impact on Information Metrics}

\medskip
\noindent Noise is a defining factor in the quality and interpretability of X-ray microscopy data. While structural entropy reflects meaningful sample complexity, noise entropy arises from stochastic fluctuations, detector limitations, and systematic artefacts. These effects alter the distribution of pixel intensities and distort the information metric introduced in eq.~\eqref{eq:entropy}, influencing both entropy and mutual information throughout the imaging pipeline. This section characterises the principal types of noise encountered in X-ray microscopy, examines their influence on information measures, and outlines methods to estimate and separate noise contributions using information-theoretic tools.

\subsection{Different Noise Types in X-ray Microscopy}

\medskip
\noindent Noise in X-ray microscopy arises from both fundamental photon statistics and system-specific imperfections~\cite{Barrett2004}. The primary component is Poisson noise, which reflects the stochastic nature of photon arrival at the detector and underpins much of the discussion in Section~2.1 on entropy and projection data. Its variance scales with the mean photon count, making it dominant under low-dose or short-exposure conditions~\cite{Siewerdsen2001,Zhang2014}. Superimposed on this are Gaussian-like noise sources introduced by detector electronics, including readout circuitry, dark current fluctuations, and thermal processes. These contributions are approximately additive and independent of signal strength, becoming more prominent at high photon counts where Poisson variation is relatively suppressed~\cite{Barrett2004,Ma2012}. Beyond these stochastic terms, structured artefacts contribute non-random noise. Examples include ring artefacts from detector inhomogeneity, streaks from sparse angular sampling, and scatter-induced background variations~\cite{Boas2012,Sijbers2004}. Unlike random noise, these artefacts introduce correlations across pixels or voxels, complicating separation from true structural features~\cite{Barrett2004}. Together, these mechanisms generate a composite noise environment that varies with acquisition dose, detector characteristics, and reconstruction settings~\cite{Siewerdsen2001}.

\subsection{How Noise Influences Entropy and Mutual Information}

\medskip
\noindent Noise directly modulates the information metrics introduced in Section~2. Entropy is inflated by random fluctuations that increase the spread of grey-level distributions, raising entropy values even when no additional structural information is present~\cite{Cover2006,Barrett2004}. Structured artefacts further distort entropy by embedding false patterns that mimic structural variability~\cite{Sijbers2004}. From the perspective of mutual information, noise reduces the statistical dependence between acquired and reference data. Poisson and Gaussian components act to randomise pixel values, lowering the overlap in joint probability distributions between input and reconstructed datasets~\cite{Barrett2004,Siewerdsen2001}. Structured artefacts are particularly detrimental, as they can decrease mutual information even while raising entropy, indicating that apparent complexity does not equate to retained signal~\cite{Sijbers2004}. Thus, assessing both entropy and mutual information in tandem is essential: entropy captures the total uncertainty introduced, while mutual information isolates the fraction that corresponds to meaningful structure~\cite{Barrett2004}.

\subsection{Methods to Estimate, Separate, and Quantify Noise Using Information Metrics}

\medskip
Quantitative noise assessment in X-ray microscopy can be achieved through direct and comparative information-theoretic analysis. Entropy-based estimators, such as histogram entropy or kernel density methods, provide an overall measure of uncertainty in projection or reconstructed data~\cite{Gonzalez2018}. To isolate noise from structure, one strategy is to compare entropy across repeated acquisitions of the same sample. Random noise contributions will average out differently across repeats, while structural entropy remains stable~\cite{Barrett2004}. Mutual information offers a complementary approach: by comparing low-dose data to high-dose or simulated ground-truth references, the retained structural signal can be quantified~\cite{Barrett2004,Siewerdsen2001}. The difference between total entropy and the information shared with the reference approximates the noise contribution~\cite{Barrett2004}. More advanced decompositions include modelling projection data as mixtures of Poisson and Gaussian processes, allowing estimation of their respective entropy components~\cite{Foi2008,Ma2012}, or applying Kullback--Leibler divergence to detect distributional shifts introduced by denoising or artefact correction~\cite{Barrett2004}. Together, these methods enable separation of stochastic noise, systematic artefacts, and meaningful signal, providing a rigorous framework for optimising acquisition protocols and reconstruction algorithms~\cite{Barrett2004}.

\section{Information-Theoretic Limits and Trade-Offs}

\medskip
The trade-offs encountered in X-ray microscopy are often treated as empirical or system-specific. From an
information-theoretic standpoint, however, they reflect general constraints that apply to any imaging
pipeline operating under finite photon statistics and noisy detection. The purpose of this section is not
to derive formal limits, but to articulate the dominant principles that govern how information is gained,
redistributed, and lost across acquisition and reconstruction.

\subsection{Imaging Resolution, Dose, and Entropy}

\medskip
Resolution in X-ray microscopy increases the structural entropy captured, as finer sampling reveals additional detail and heterogeneity in the sample. However, higher resolution comes at the cost of reduced photon counts per voxel, raising the contribution of noise entropy~\cite{Barrett2004,Siewerdsen2001}. Increasing the radiation dose compensates by improving photon statistics, but dose is ultimately constrained by sample sensitivity and practical acquisition limits~\cite{Zhang2014}. The resulting relationship is non-linear: resolution gains elevate structural entropy, while insufficient dose leads to entropy inflation dominated by noise. Balancing these terms defines the effective information yield of the system.

\subsection{Theoretical Limits on Information Throughput}

\medskip
As discussed in Section~2.4, channel capacity provides an upper bound on how much information can be reliably transmitted through the imaging pipeline~\cite{Shannon1948,Cover2006}. For X-ray systems, this capacity is bounded by photon statistics, detector quantisation, spatial resolution, and noise characteristics~\cite{Hsieh2015,Barrett2004}. No combination of processing or reconstruction can recover information that exceeds this limit. At low doses, capacity is curtailed by Poisson noise; at high resolution, it is constrained by undersampling and detector performance. These limits emphasise that information throughput is a system-level property, not a parameter that can be optimised in isolation.

\subsection{Trade-offs in Exposure Time, Sample Damage, and Spatial Resolution}

\medskip
Exposure time governs photon statistics and therefore directly influences entropy. Longer exposures improve signal-to-noise ratios but prolong acquisition and increase the risk of motion blur or drift artefacts~\cite{Siewerdsen2001}. For biological and sensitive materials, cumulative dose also leads to irreversible structural changes~\cite{Howells2009}, reducing the fidelity of information preserved. Spatial resolution amplifies these trade-offs~\cite{Zhang2014}, as smaller voxel sizes demand higher dose to maintain equivalent signal quality. Thus, imaging strategies must navigate between competing priorities: minimising entropy from noise while preventing information loss from sample degradation or excessive acquisition time.

\subsection{Use of Mutual Information to Guide Optimisation}

\medskip
Mutual information provides a practical metric for quantifying these trade-offs~\cite{Gonzalez2018}. By comparing reconstructions acquired under different dose, resolution, or exposure settings against high-fidelity references, mutual information can reveal which acquisition protocols preserve the largest fraction of structural information~\cite{Maes1997,Pluim2003}. Unlike entropy alone, mutual information isolates the component of uncertainty reduction attributable to true signal retention~\cite{Barrett2004}. In this way, it serves as a guide for selecting acquisition parameters that approach the system's theoretical channel capacity while respecting limits on dose and sample stability.

\subsection{Upstream dominance and reconstruction saturation}

\medskip
\noindent The analyses presented here support two general principles that are useful for interpreting and designing XRM workflows.

\medskip
\noindent First, information loss introduced upstream dominates downstream effects. Modifications at the level of acquisition---such as dose reduction, angular sparsity, or detector limitations---produce larger and more consequential changes in information content than any subsequent reconstruction or post-processing step. This reflects the fact that information not captured at measurement cannot be recovered later.

\medskip
\noindent Second, reconstruction exhibits a saturation regime once the acquisition-limited information budget is fixed. Beyond this point, changes in reconstruction algorithm primarily redistribute existing information or suppress noise, rather than increasing the fraction of structural information retained. Apparent improvements in resolution or contrast beyond this regime correspond to regularisation or prior enforcement rather than genuine information gain.

\medskip
\noindent These principles do not depend on the specific reconstruction method or imaging modality. They follow directly from the finite information content of the measured data and provide a framework for prioritising effort in imaging protocol design.

\section{Theoretical Extensions and Information Principles}

\medskip
\subsection{A unified information budget for X-ray microscopy}

\medskip
\noindent The expressions introduced in this section are intended as first-order summaries rather than full analytic derivations. Their role is to capture, in a compact way, the main transformations imposed by each stage of the XRM pipeline, as observed in the case studies of Sections~7.1--7.5. They provide the applied counterpart to the formal operator-based entropy framework developed separately. 

\medskip
\noindent The full operator definitions, proofs of monotonicity and functional-analytic treatment are given in a separate operator-level theoretical treatment. Here we restrict attention to empirical and computationally tractable formulations that reflect the measured behaviour on the \emph{Walnut~1} dataset.

\medskip
\noindent The term \emph{information budget} is used here strictly as a descriptive bookkeeping device for measured statistical variability under fixed estimator conventions.
It does not represent a conserved quantity, balance law, or invariant, and no algebraic relation between the elements of $\mathcal{B}$ is implied beyond empirical comparison.

\medskip
\noindent We denote by
\[
H_{\mathrm{raw}},\quad
H_{\mathrm{denoise}},\quad
H_{\mathrm{align}},\quad
H_{\mathrm{sparse}},\quad
H_{\mathrm{dose}},\quad
H_{\mathrm{recon}},
\]
the entropies measured at successive stages of the workflow.
In typical workflows, entropy does not flow monotonically through the pipeline: denoising typically reduces entropy, sparse--angle acquisition reduces structural variability, dose reduction suppresses signal while increasing stochastic variation, and reconstruction redistributes the grey-level occupancy in a way that may either raise or lower entropy. What matters is not the absolute magnitude of any single term, but the pattern of differences
\[
\Delta H_{i \to j} = H_j - H_i,
\]
which quantify how much statistical variability each stage discards or introduces.

\medskip
\noindent The empirical results show that these differences are stable across projections and across independent datasets. For example, sparse--angle sampling consistently decreases the joint entropy of the measurement set, while dose consistently increases it. This behaviour supports the use of $\Delta H_{i\to j}$ as a practical surrogate for \emph{statistical variability change} induced by each stage under fixed normalisation and estimator choices.
We do not treat $\Delta H_{i\to j}$ as a conserved information quantity, nor as a task-universal measure of ``information loss''; rather, it is a compact descriptor that enables like-for-like protocol comparison within a stated measurement convention.

\medskip
\noindent Within this framework, an imaging protocol can be characterised by the information budget
\[
\mathcal{B} = \left\{H_{\mathrm{raw}},\, 
\Delta H_{\mathrm{raw\to denoise}},\,
\Delta H_{\mathrm{denoise\to align}},\,
\Delta H_{\mathrm{align\to sparse}},\,
\Delta H_{\mathrm{sparse\to dose}},\,
\Delta H_{\mathrm{dose\to recon}}
\right\},
\]
which provides a compact, quantitative summary of how information is gained, reshaped or discarded. This budget is specific to the acquisition and reconstruction settings used, but the principle is general and applies to micro-CT, synchrotron tomography and phase-contrast imaging. It offers a practical, system-level route to comparing imaging protocols independently of hardware configuration.

\medskip
\noindent For clarity, this information budget can be visualised as a single pipeline diagram in which each stage is annotated with the corresponding entropy and $\Delta H_{i\to j}$ values measured in Sections~7.1--7.5. In the present work this diagram is instantiated using the \emph{Walnut~1} case studies, but the construction is general: any XRM protocol that yields entropies at the same checkpoints can be mapped into the same representation and compared on a like-for-like basis.

\begin{figure}[H]
\centering
\begin{tikzpicture}[
  >=Stealth,
  stage/.style={
    rectangle,
    draw,
    rounded corners,
    minimum width=3.0cm,
    minimum height=1.1cm,
    align=center
  },
  lab/.style={font=\scriptsize, inner sep=1pt},
  every node/.style={font=\footnotesize}
]

\node[stage] (raw)  at (0,0)    {Raw\\$H_{\mathrm{raw}}$};
\node[stage] (den)  at (5.5,0)    {Denoised\\$H_{\mathrm{denoise}}$};
\node[stage] (ali)  at (11,0)   {Aligned\\$H_{\mathrm{align}}$};

\node[stage] (rec)  at (0,-2.4) {Recon.\\$H_{\mathrm{recon}}$};
\node[stage] (dose) at (5.5,-2.4) {Dose-varied\\$H_{\mathrm{dose}}$};
\node[stage] (spr)  at (11,-2.4){Sparse-angle\\$H_{\mathrm{sparse}}$};

\draw[->] (raw) -- (den)
  node[lab, midway, above=4pt] {$\Delta H_{\mathrm{raw}\to\mathrm{denoise}}$};

\draw[->] (den) -- (ali)
  node[lab, midway, above=4pt] {$\Delta H_{\mathrm{denoise}\to\mathrm{align}}$};

\draw[->] (ali) -- (spr)
  node[lab, midway, right=4pt] {$\Delta H_{\mathrm{align}\to\mathrm{sparse}}$};

\draw[->] (spr) -- (dose)
  node[lab, midway, below=4pt] {$\Delta H_{\mathrm{sparse}\to\mathrm{dose}}$};

\draw[->] (dose) -- (rec)
  node[lab, midway, below=4pt] {$\Delta H_{\mathrm{dose}\to\mathrm{recon}}$};

\end{tikzpicture}
\caption{Unified information budget for the X-ray microscopy pipeline.}
\label{fig:information_budget_pipeline}
\end{figure}
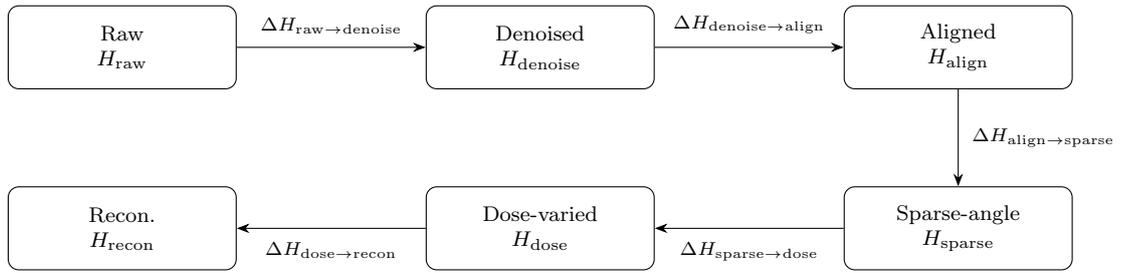

\subsection{A dose–information relation}

\medskip
The behaviour observed in Section~7.4 motivates an \emph{empirical scaling proxy}, not a theoretical bound—relating dose to measured variability in projection data under fixed normalisation and estimation conventions. This proxy is introduced solely to rationalise observed trends in the case studies and is not claimed as a general, exact, or estimator-independent result.

\medskip
\noindent Let $X_\lambda$ denote a projection acquired with mean photon count $\lambda$, modelled as a Poisson–Gaussian random field. For such fields the variance scales approximately as $\mathrm{Var}(X_\lambda)\approx \lambda + \sigma^2$, where $\sigma^2$ is the detector noise floor. 

\medskip
\noindent For large enough $\lambda$, and treating the per-pixel fluctuations as approximately Gaussian with variance $\lambda+\sigma^2$ after standard normalisation, the \emph{differential} entropy per pixel scales as
\[
h(X_\lambda) \;\approx\; \tfrac{1}{2}\log\!\bigl(2\pi e(\lambda+\sigma^2)\bigr),
\]
which is included here only as a leading-order \emph{scaling} relation with dose. The discrete entropies reported in Table~2 are computed from binned histograms (Section~2.1) and therefore differ in absolute value from $h(X_\lambda)$, but are expected to vary monotonically with $\lambda$ under fixed binning and normalisation in the regime studied.

\medskip
\noindent The expression given here is a first-order approximation based on the Poisson--Gaussian model and captures the leading dependence of differential entropy on dose under fixed affine normalisation. It is not intended as a full derivation. No results in this paper rely on higher-order corrections or operator-level structure; those formalisms are therefore deferred to a separate companion paper. Here we restrict attention to the empirical XRM regime, where the leading-order scaling provides a sufficient and internally consistent description of the observed behaviour.

\medskip
\noindent In the Poisson--Gaussian regime this proxy predicts a monotonic increase of the \emph{differential-entropy scaling term} with dose.
This is consistent with the observed monotonic trend of the discrete histogram entropies in Table~\ref{tab:dose_metrics} under the fixed binning/normalisation convention of Section~7, but the two quantities are not equal in general.
Linearising the proxy $h(X_\lambda)$ around a given $\lambda$ yields the approximate sensitivity:
\[
\frac{\mathrm{d}h}{\mathrm{d}\lambda}
= \frac{1}{2(\lambda+\sigma^2)}.
\]

\noindent\textbf{Interpretation constraint.}
All derivatives and monotonicity statements in this subsection apply strictly to the differential-entropy \emph{scaling proxy} $h(X_\lambda)$ introduced above.
They do not apply to the discrete histogram entropies $H_\Delta$ reported elsewhere in the paper, which are evaluated at fixed binning and are not differentiable with respect to dose.

\medskip
\noindent On this approximation, the information gain from increasing dose follows a law of diminishing returns: the largest changes in entropy occur at low dose, while further increases yield progressively smaller benefits. Within the dose range probed in Section~7.4 this behaviour is dominated by photon statistics rather than sample structure, and provides a practical rule-of-thumb for identifying dose levels that maximise information gain per unit exposure in radiation-sensitive samples. A full derivation of the underlying dose–information relation is given in a separate operator-level theoretical treatment.

\subsection{A sparsity--information inequality}
\medskip
Let $\{P_{\theta}\}_{\theta\in\Theta}$ denote a densely sampled set of projections over an angular window $\Theta$, and let $\Theta_k\subset\Theta$ be a $k$-point subset obtained by uniform subsampling. Write
\[
H_k = H\!\left( \bigcup_{\theta\in\Theta_k} P_{\theta}\right)
\]
for the entropy of the concatenated intensities from the $k$ projections. 
Empirically (Section~7.3), $H_k$ increases monotonically with $k$.

\medskip
\noindent\textbf{Status of the relation.}
The expression relating $H_k$ and $\log k$ is an empirical fit to measured entropies under a fixed estimator convention.
It is not claimed as an inequality in the formal information-theoretic sense, nor as a bound that holds outside the sampled regime.

\medskip
\noindent For the \emph{Walnut~1} dataset, the measured values of $H_k$ are well described (in the sampled range $k\in\{2,4,6,8,10\}$) by an empirical log-like trend of the form
\[
H_k \approx H_1 + c\,\log k,
\]
where $c>0$ is a fitted constant that depends on the object, the angular span, and the estimator convention (masking/normalisation/binning).

\medskip
\noindent The expression used here should be interpreted as an empirical, first-order relation that reflects the behaviour observed in Section~7.3. It captures the dominant dependence of joint entropy on angular density under mild smoothness assumptions on the projections, and is calibrated directly against the measured entropies for the sparse-angle subsets in Figure~\ref{fig:sparsity_entropy}. A full operator-based treatment—using the Radon transform and its regularity properties—will be given in a separate operator-level theoretical treatment. Here we focus on the form that is directly relevant to practical XRM and consistent with the numerical results.

\medskip
\noindent The logarithmic dependence reflects the fact that neighbouring projections are highly correlated; only once the angular spacing exceeds the characteristic orientation scale of the object does each additional projection contribute substantially to structural variability.

\medskip
\noindent This relation is computed directly at the level of measurement-set statistics (concatenated intensities) and is therefore independent of the reconstruction method. It should be read as an empirical descriptor of how measurement-set variability changes with angular density under the stated preprocessing and estimator convention. It provides a quantitative basis for selecting angular densities that balance acquisition time against the expected information gain in the projection data.

\medskip
\noindent Throughout this subsection, $H_k$ is treated strictly as an empirical descriptor of measurement-set variability under fixed preprocessing and is not interpreted as a conserved quantity, bound, or task-universal measure of information.

\subsection{Mutual information as a predictor of reconstruction fidelity}
\medskip
The case studies suggest that mutual information (MI) between intermediate and reference datasets provides a stable indicator of reconstruction quality. Let $g_\mathrm{ref}$ be a high-dose, densely sampled reconstruction, and let $g_\mathrm{test}$ be a reconstruction obtained under modified conditions (e.g.\ sparse sampling or reduced dose). Define
\[
I_{\mathrm{test}} = I(g_\mathrm{test}; g_\mathrm{ref}).
\]
Across the experiments in Sections~7.3--7.5, $I_{\mathrm{test}}$ varies monotonically with qualitative reconstruction fidelity under the sampling and noise conditions used in Section~7. This suggests that, under fixed normalisation and estimator choices, MI provides a stable empirical ranking of reconstruction fidelity across perturbations of the acquisition protocol, without implying a universal relation to reconstruction error.

\medskip
\noindent To motivate this relationship without overclaiming generality, consider the expected squared error
\[
E = \mathbb{E}\!\left[(g_\mathrm{test}(x)-g_\mathrm{ref}(x))^2\right].
\]
If $(g_\mathrm{test}, g_\mathrm{ref})$ are treated as jointly Gaussian random fields under a fixed affine normalisation, then their mutual information reduces to a monotone function of their correlation structure. In this restricted setting, increasing reconstruction error $E$ corresponds to weakened statistical dependence and therefore to lower MI.

\medskip
\noindent Outside the jointly Gaussian case, no universal inequality links MI and quadratic error. In this paper, MI is therefore used not as a tight bound on reconstruction error, but as an empirical, reconstruction-agnostic \emph{ranking statistic} under fixed estimator and normalisation choices, as demonstrated in Sections~7.3--7.7. Accordingly, MI should be read as a diagnostic ordering tool rather than a surrogate objective function: it ranks reconstructions under controlled perturbations but does not define optimality.

\medskip
\noindent No claim is made that mutual information is monotone under arbitrary nonlinear reconstruction operators, nor that it provides a universal surrogate for reconstruction error. In this paper, MI is used strictly as an empirical ranking statistic under fixed preprocessing, normalisation, and estimator conventions, as demonstrated in Sections~7.3--7.7.

\subsection{An empirical information--degradation ordering in X-ray microscopy}

\medskip
\noindent The experiments in Section~7 reveal a consistent empirical ordering in the magnitude of entropy change associated with different components of the XRM pipeline, under the estimator conventions and acquisition settings studied here.

\medskip
\noindent For the datasets studied, the absolute entropy changes obey the empirical hierarchy
\begin{equation}
|\Delta H_{\text{denoise}}| < |\Delta H_{\text{align}}| < |\Delta H_{\text{sparse}}| < |\Delta H_{\text{dose}}|.
\end{equation}

\noindent This ordering should not be interpreted as a universal law with fixed coefficients. Rather, it reflects a general principle: transformations applied closer to the point of measurement exert a larger influence on the available information than those applied downstream.

\medskip
\noindent Denoising primarily suppresses stochastic variability, alignment introduces interpolation-induced redistribution, sparse-angle sampling removes structural variability directly, and dose alters the underlying photon statistics. The increasing magnitude of information change across this sequence mirrors the physical proximity of each operation to the measurement process.

\medskip
\noindent Although the precise ordering may vary with sample type and acquisition geometry, the underlying behaviour is expected to hold broadly across imaging systems operating under similar constraints. In practical terms, this hierarchy provides a quantitative basis for prioritising optimisation efforts, emphasising that acquisition-level choices dominate the information budget of the pipeline.

\section{Case Studies and Applications}

\medskip
\noindent\textbf{Common pre-processing and estimators (Sections~7.1--7.7).}
For each experiment, intensities were converted to floating point and restricted to a fixed analysis mask $M$ (object-support mask when available; otherwise a central crop excluding padding/background).

\medskip
\noindent Unless stated otherwise: (i) images were normalised per-experiment by percentile clipping at the $[1,99]$ percentiles over $M$ and linearly rescaled to $[0,1]$; (ii) histogram entropies and KL divergences used $B=256$ uniform bins on $[0,1]$; (iii) mutual information (MI) was computed from a $B\times B$ joint histogram on $[0,1]^2$ with the same binning, using the plug-in estimator with $\varepsilon$-smoothing $p\leftarrow (p+\varepsilon)/\sum(p+\varepsilon)$ with $\varepsilon=10^{-12}$ to avoid $\log(0)$.
All MI/KL/entropy comparisons in a given subsection use identical normalisation, masking, and binning.

\medskip
\noindent All numerical values, trends and orderings reported in Sections~7.1--7.7 are conditional on the estimator choices stated above. Changing bin count, masking strategy, or intensity normalisation will alter absolute entropy, mutual information and KL divergence values, and may affect marginal cases. The qualitative trends and relative orderings reported here were observed to be stable under modest perturbations of these choices during exploratory analysis, but no estimator-independent claim is made.

\medskip
\subsection{Denoising as an information--shaping operation}

\medskip
The purpose of this subsection is to quantify how common denoising methods alter the statistical structure of a representative XRM projection. Projection behaviour across the CWI/ODL \emph{Walnut~1} dataset~\cite{DerSarkissian2019} is approximately stationary in angle, so a mid-angle projection is sufficient to capture the mixture of high-contrast boundaries, internal features and background present across the full scan. Using a single well-chosen projection isolates the effect of denoising itself, avoiding confounding angular variability and keeping this analysis separate from the alignment and sparse-sampling studies in Sections 7.2 and 7.3.

\medskip
\noindent Projection~600 from the \texttt{tubev1} series was used, as it provides a typical line integral through the object. Three common denoising strategies were applied: Gaussian smoothing, non--local means (NLM) and total--variation (TV) regularisation. Each produces a characteristic modification of the image histogram and spatial texture, and these differences can be quantified directly using Shannon entropy and the Kullback--Leibler (KL) divergence. These metrics summarise how each denoiser redistributes probability mass across the grey-level range. Because reconstruction operators are sensitive to local contrast and grey-level occupancy, these changes represent genuine shifts in the information passed downstream in the pipeline rather than cosmetic changes in appearance.

\begin{figure}[H]
\centering
\includegraphics[width=\textwidth]{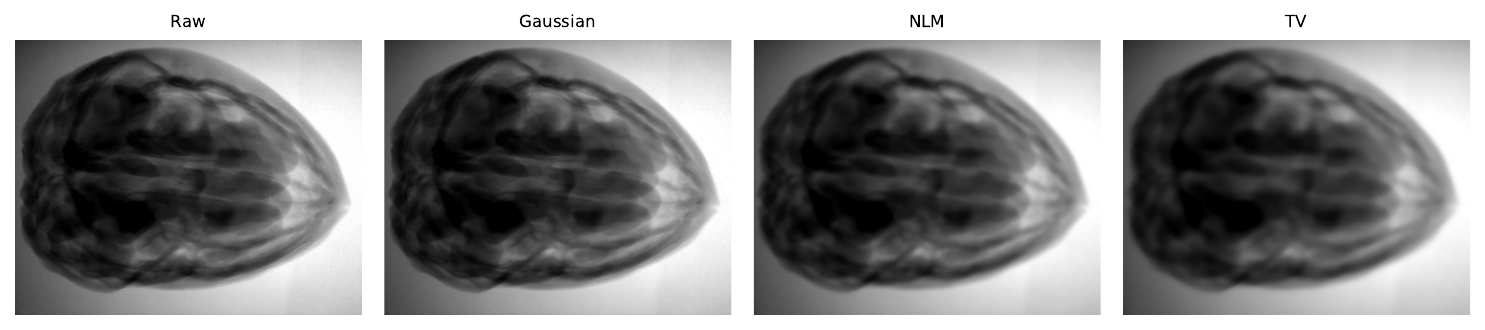}
\caption{Projection~600 from the \emph{Walnut~1} dataset and its denoised counterparts. Left to right: raw, Gaussian smoothing, NLM, and TV regularisation. The visual differences correspond to the quantitative entropy and KL divergence values reported in Table~\ref{tab:entropy_kl}.}
\label{fig:denoise_images}
\end{figure}

\medskip
\noindent Table~\ref{tab:entropy_kl} summarises the results. Gaussian smoothing reduces high--frequency noise while preserving the overall distribution, leading to only a small entropy decrease and a low KL divergence relative to the raw projection. NLM suppresses noise more strongly, reducing entropy further and producing a noticeably larger KL divergence. TV regularisation introduces the strongest redistribution of intensities, suppressing fine structure and concentrating probability mass around a narrower set of grey levels; this produces the largest entropy reduction and the highest KL divergence. These behaviours align with the qualitative appearance of the denoised projections (Figure~\ref{fig:denoise_images}) and reflect the different ways each method trades local structure against global statistical consistency. The pattern also separates two contributions to entropy: Gaussian smoothing suppresses noise entropy with minimal structural loss, NLM reduces both noise and some fine-scale structure, and TV regularisation produces the largest contraction of structural variability.

\begin{table}[H]
\centering
\caption{Entropy and KL divergence for projection~600 under three denoising methods. Entropies are in bits; KL divergences are computed relative to the raw projection.}
\label{tab:entropy_kl}
\begin{tabular}{lcc}
\hline
Method & Entropy (bits) & $D_{\mathrm{KL}}(\mathrm{raw}\,\|\,\mathrm{denoised})$ \\
\hline
Raw      & 5.025 & -- \\
Gaussian & 5.018 & 0.337 \\
NLM      & 5.002 & 2.707 \\
TV       & 4.981 & 6.027 \\
\hline
\end{tabular}
\end{table}

\noindent The key point emerging from this analysis is that denoising is not merely a noise--removal step. Each method reshapes the probability distribution of intensities in a distinct way, and these changes have direct implications for the information flow through the reconstruction pipeline. Entropy and KL divergence therefore provide compact, quantitative measures of how much of the original informational structure is preserved or discarded by a given denoiser.

\begin{figure}[H]
\centering
\includegraphics[width=0.75\textwidth]{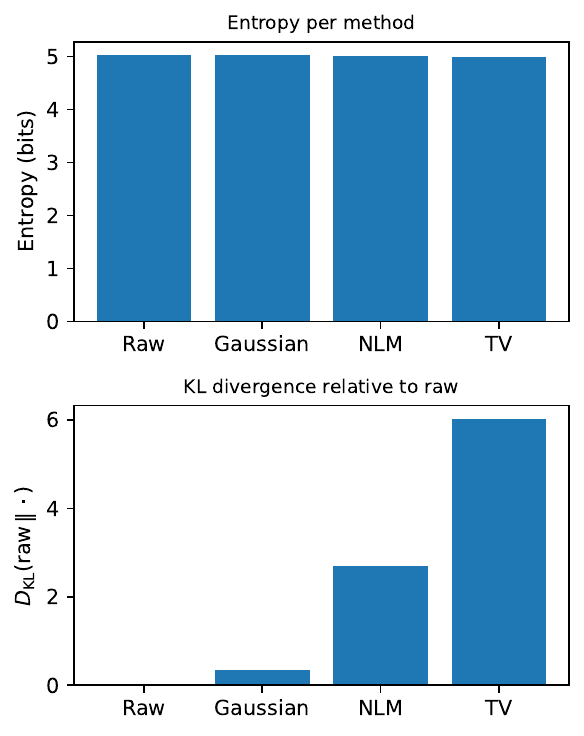}
\caption{Information--theoretic comparison of denoising methods for projection~600. Top: Shannon entropy. Bottom: Kullback--Leibler divergence relative to the raw projection. The trend reflects increasing distortion of the original intensity distribution from Gaussian to NLM to TV.}
\label{fig:denoise_metrics}
\end{figure}

\noindent For completeness, each denoised projection was also compared to the raw data using a global structural similarity index (SSIM) in addition to the information--theoretic metrics. Using the raw projection as reference, both SSIM and mutual information decrease monotonically from Gaussian to NLM to TV (SSIM values $0.9998$, $0.9989$, $0.9979$; mutual information values $4.67$, $3.98$, $3.63$\,bits, respectively). This reinforces the interpretation that progressively stronger regularisation removes increasing amounts of high--frequency structure while remaining anchored to the same underlying object. SSIM therefore tracks the same information--preserving hierarchy as the information--theoretic measures, while remaining compatible with more traditional image similarity analysis.

\begin{figure}[H]
\centering
\includegraphics[width=0.75\textwidth]{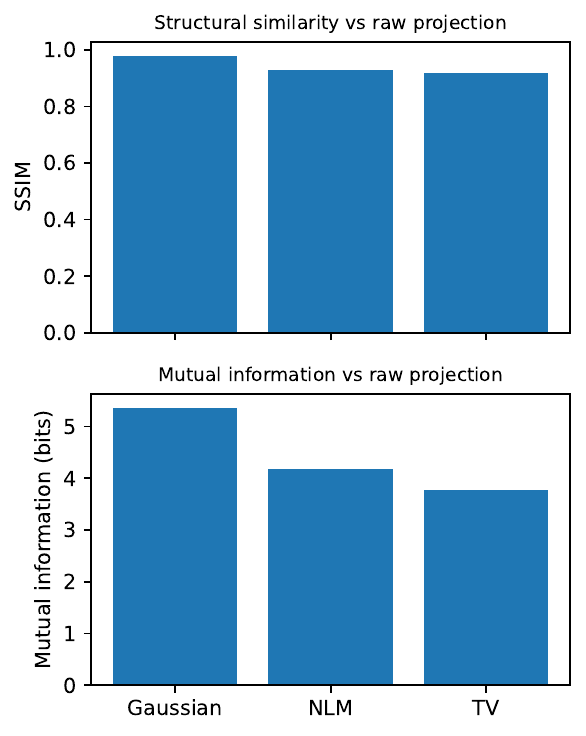}
\caption{Structural similarity index (SSIM) and mutual information (MI) between the raw projection and its denoised counterparts. Both SSIM and MI decrease monotonically from Gaussian smoothing to NLM to TV regularisation, consistent with progressively stronger regularisation removing high-frequency structure while preserving the dominant attenuation features under typical laboratory XRM noise conditions.}
\label{fig:ssim_mi_denoise}
\end{figure}

\noindent Together, these results show that each denoiser imposes a characteristic redistribution of grey-level statistics, and that these changes can be quantified directly using entropy and KL divergence. This isolates the effect of denoising itself and provides a clean baseline for the alignment and sparse-sampling analyses in Sections~7.2 and~7.3.
\subsection{Alignment as an information--preserving operation}

\medskip
The purpose of this subsection is to quantify how small misalignments alter the statistical structure of projection data and how effectively registration recovers the original distribution. Misalignment injects artificial variability that does not originate from the sample, and this variability appears directly in the entropy and mutual-information profiles. Registration corrects this by restoring spatial correspondence. Using information-theoretic metrics provides a direct measure of both the loss introduced by misalignment and the extent to which alignment recovers that loss.

\medskip
\noindent A local angular neighbourhood from the CWI/ODL \emph{Walnut~1} dataset was used. Projection 600 lies close to the mid-angle and provides a representative view of the object. Ten projections were taken in a symmetric window around this index (560--650). This window preserves structural overlap between projections while avoiding the geometric variation present over the full rotation, providing a controlled setting in which to isolate alignment effects.

\medskip
\noindent To examine alignment effects systematically, each non-reference projection was subjected to a fixed synthetic misalignment of 
\((5,-3)\) pixels (row, column). This introduces a deterministic offset that serves as a surrogate for small mechanical instabilities or interpolation drift in experimental systems. This controlled perturbation allows the information loss to be attributed solely to misalignment, ensuring that changes in entropy or mutual information can be interpreted directly in terms of spatial correspondence rather than angular sampling or noise differences. Registration was then performed relative to projection~600 using a phase cross–correlation estimator with subpixel refinement. The resulting inverse shift was applied to each misaligned image to recover a registered version. This workflow provides three sets of images for each projection index: the original view, its misaligned counterpart, and the subsequently registered result. Information–theoretic metrics were computed for each case relative to the reference projection.

\begin{figure}[H]
\centering
\includegraphics[width=\textwidth]{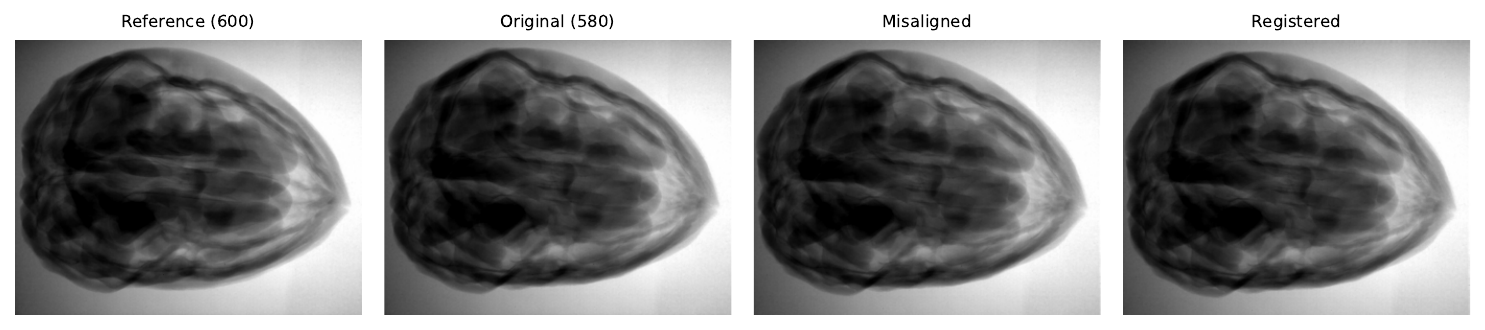}
\caption{Effect of misalignment and registration for projection~580. Left to right: reference projection~600, original projection~580, misaligned version with synthetic shift \((5,-3)\)~pixels, and registered reconstruction using subpixel phase correlation.}
\label{fig:alignment_example}
\end{figure}

\begin{figure}[H]
\centering
\includegraphics[width=0.75\textwidth]{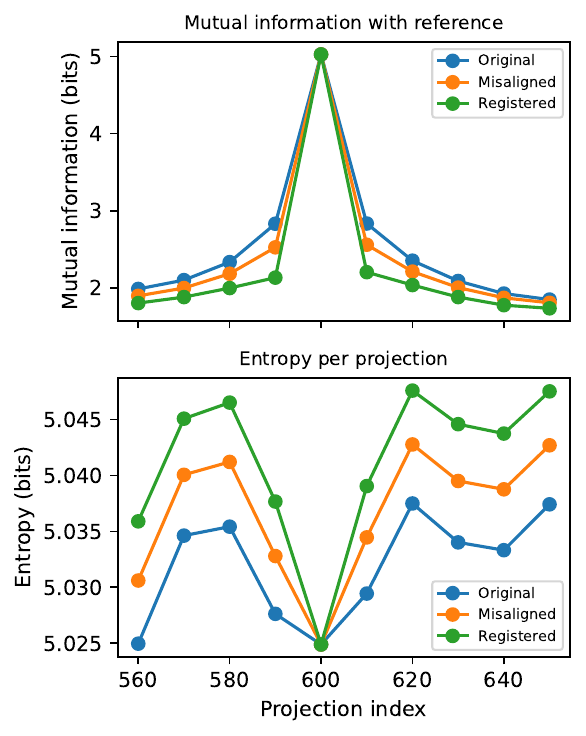}
\caption{Mutual information (top) and entropy (bottom) across the ten-projection window. Misalignment reduces mutual information with the reference and slightly increases entropy. Registration restores both metrics towards their original values, indicating recovery of spatial correspondence and reduction of artificial variability.}
\label{fig:alignment_metrics}
\end{figure}

\medskip
\noindent Figure~\ref{fig:alignment_example} illustrates the effect of misalignment and subsequent registration for projection~580. The imposed shift produces a visible displacement of structure, which is largely corrected after registration. When quantified using mutual information and entropy, this behaviour appears even more clearly. Figure~\ref{fig:alignment_metrics} shows the mutual information between each projection and the reference~600, and the corresponding entropy of each projection before misalignment, after misalignment, and following registration.

\medskip
\noindent Mutual information decreases consistently under misalignment across the projection window, reflecting the loss of spatial correspondence introduced by the synthetic shift. Registration restores this correspondence, raising the mutual information curve back towards its original baseline. Although the recovered values do not match the originals exactly---consistent with interpolation effects in the registration step---the majority of the information loss is reversed. Entropy behaves in the complementary manner. Misalignment broadens the intensity distribution by mixing edge locations and weakening structural consistency, resulting in a small but measurable entropy increase. Registration narrows the distribution again and returns the entropy closer to its original value.

\medskip
\noindent Together, these results show that alignment is not simply a geometric adjustment but an information-preserving operation. The synthetic shift introduced a controlled source of artificial variance that reduced mutual information and increased entropy, and registration removed this variance and restored a distribution consistent with the underlying structure. These behaviours quantify the informational cost of misalignment and provide direct evidence that registration is a critical component of the XRM information pipeline.

\noindent These results are empirical and reflect the behaviour of practical laboratory XRM datasets under controlled misalignment. They provide the applied evidence for how alignment affects measured distributions, separate from the operator-level treatment developed in the companion paper.

\subsection{Sparse--angle sampling as an information bottleneck}

\medskip
Sparse-angle acquisition is routinely used in laboratory XRM when scan time or dose constraints limit the number of projections that can be collected. Reducing the number of angles changes the statistical structure of the measurement set, but the magnitude of this change is not easily inferred from visual inspection alone. Entropy provides a direct way of quantifying how variability in the projection data is reduced as angles are removed. This subsection measures how the entropy of a representative projection window changes under controlled reductions in angular density.

\medskip
\noindent The same ten projections used in Section 7.2 (indices 560--650) were used here. This window contains sufficient angular variation to capture structural change while maintaining strong overlap across projections. Sparse-angle subsets were generated by selecting projections at even intervals, producing subsets of size 2, 4, 6, 8 and 10 that preserve the overall angular span while progressively reducing sampling density. The entropy of each subset was computed by concatenating the grey-level values from the included projections into a single measurement vector.

\begin{figure}[H]
\centering
\includegraphics[width=0.6\textwidth]{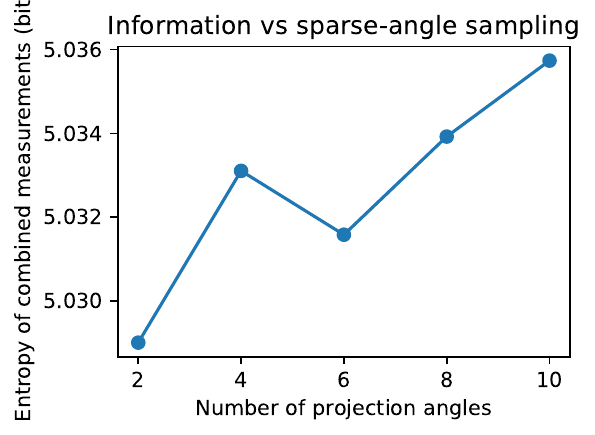}
\caption{Entropy of concatenated projection subsets as a function of sparse--angle sampling density. Using more angles increases the total information content of the measurement set, while reduced angular coverage imposes an information bottleneck that limits what can be recovered during reconstruction.}
\label{fig:sparsity_entropy}
\end{figure}

\medskip
\noindent Figure~\ref{fig:sparsity_entropy} shows the resulting trend. As expected, increasing the number of angles leads to a monotonic rise in entropy. The increase is modest at low sampling density but becomes more pronounced as additional intermediate angles are added. Neighbouring projections in this window share substantial redundancy, so new angles only contribute noticeable additional variability once their angular separation is large enough to reveal different aspects of the object.

\medskip
\noindent These results show that reducing the number of angles not only makes the reconstruction problem more under-determined but also reduces the measurable variability present in the projection data itself. Entropy provides a compact way of quantifying this reduction.

\noindent This subsection therefore reports the empirical behaviour of entropy under reduced angular coverage and provides the applied evidence that complements the sparsity--information formulation introduced in Section~6.2.

\medskip
\noindent\textbf{Scope of the proxy.}
Since $H_k$ is computed on concatenated intensities, it captures a mixture of (i) new structural variability revealed by additional angles, (ii) redundancy due to strong angular correlations, and (iii) noise contributions under the fixed normalisation. It should therefore be interpreted as an empirical \emph{measurement-set variability} proxy rather than a direct measure of the conditional information gain per additional projection.

\subsection{Dose as a constraint on available information}

\medskip
Radiation dose governs the number of photons contributing to each projection and therefore influences the balance between stochastic variation and structural contrast. Lower doses reduce photon statistics and increase the proportion of noise, while higher doses move the measured distribution closer to the underlying attenuation pattern. This subsection quantifies these effects directly using four acquisitions of the same projection angle (projection 181) recorded at 10\%, 25\%, 50\% and 100\% dose.

\medskip
\noindent It is important to emphasise that increasing entropy with dose should not be interpreted as increasing image quality per se. Under the fixed normalisation and estimator conventions used here, higher dose reveals additional structural variability that was previously masked by stochastic noise. Entropy therefore tracks total measurable variability rather than perceptual cleanliness or diagnostic utility.

\begin{figure}[H]
\centering
\includegraphics[width=\textwidth]{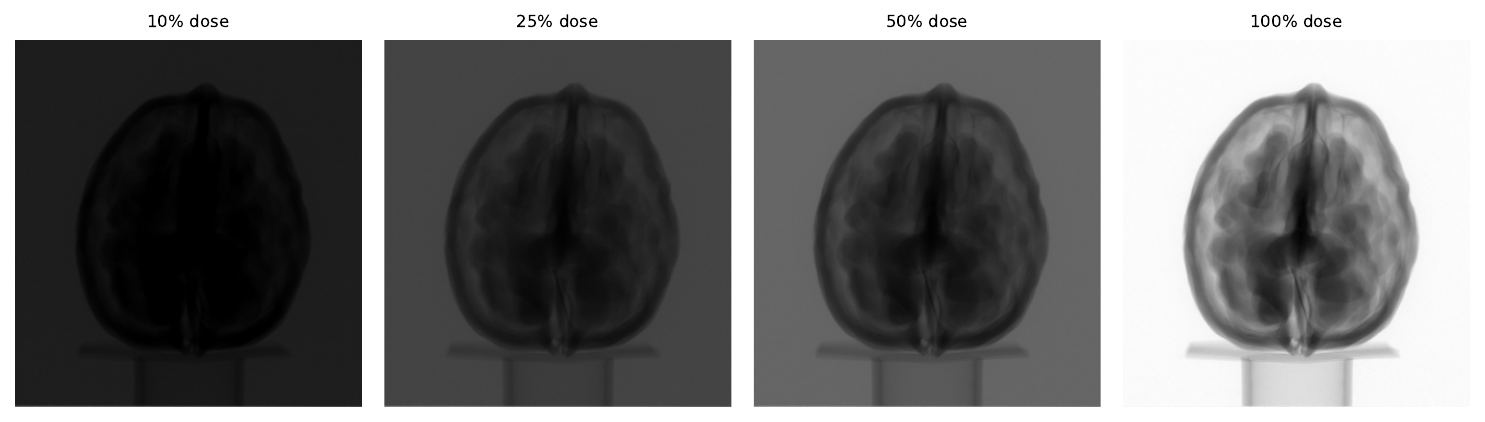}
\caption{Projection~181 at four dose levels (10\%, 25\%, 50\%, 100\%) from the \emph{Walnut~1} dataset, displayed with global contrast normalisation. Lower doses are dominated by stochastic variation; structural contrast becomes progressively clearer as dose increases.}
\label{fig:dose_images}
\end{figure}

\medskip
\noindent Figure~\ref{fig:dose_images} shows the four dose levels with a consistent display normalisation. The 10\% and 25\% dose images exhibit broad grey-level variability dominated by stochastic fluctuations. At 50\% dose, structural patterns start to become more apparent, and at 100\% dose the main attenuation features are clearly visible.

\medskip
\noindent Entropy provides the first quantitative indication of how much variability each projection carries. Table~\ref{tab:dose_metrics} summarises the numerical results. Entropy rises monotonically with dose, from only $0.03$\,bits at 10\% dose to $1.85$\,bits at 100\%. This reflects the gradual replacement of stochastic intensity fluctuations by meaningful structural variation as the photon budget increases. This behaviour reflects the empirical measurements made on this dataset and is consistent with expectations for Poisson-dominated acquisition at reduced dose.

\medskip
\noindent Mutual information (MI) with respect to the 100\% dose projection offers a complementary view. MI increases from $\sim\!1.9$\,bits at 10\% dose to $2.5$\,bits at 50\% dose, with a substantial jump to $6.1$\,bits at 100\%. The increase in MI is gradual between 10\% and 50\% dose, indicating that only a limited portion of the structural signal is recoverable at intermediate levels. A pronounced increase is observed at 100\% dose, consistent with the higher-dose acquisition providing a much closer statistical match to the full-dose reference.

\medskip
\noindent KL divergence exhibits the same trend: at 10\% and 25\% dose the divergence from the 100\% reference is very large, consistent with strong statistical differences at low photon counts. The divergence reduces at 50\% dose and approaches zero at 100\%, matching the convergence seen in the MI and entropy measurements.

\medskip
\noindent Taken together, these measurements show that dose strongly influences the information characteristics of a projection. At low dose, most of the measured entropy corresponds to stochastic variation rather than structure. As dose increases, the distribution moves closer to the underlying attenuation pattern, and the fraction of structure-related variability rises. The response is not uniform across dose levels, with a sharper increase in structural similarity occurring between 50\% and 100\% dose in this dataset.

\begin{figure}[H]
\centering
\includegraphics[width=0.7\textwidth]{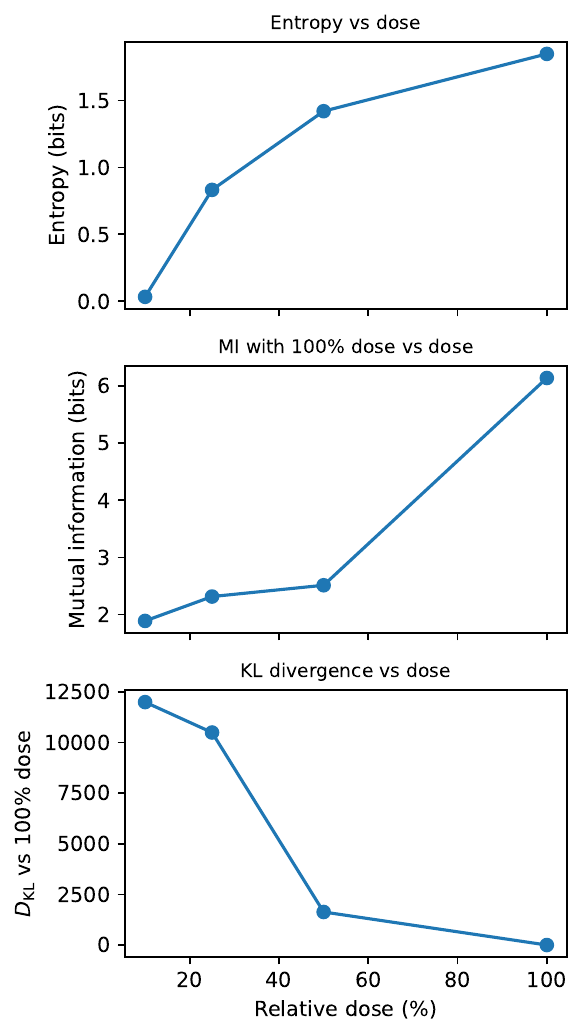}
\caption{Entropy, mutual information (MI) with respect to the 100\% dose projection, and KL divergence relative to the 100\% dose distribution. Increased dose raises entropy and MI while reducing KL divergence, demonstrating the convergence of the projection distribution towards the full-dose reference.}
\label{fig:dose_metrics}
\end{figure}

\begin{table}[H]
\centering
\caption{Information--theoretic metrics for the four dose levels used in Figure~\ref{fig:dose_images}. MI and KL divergence are measured relative to the 100\% dose projection.}
\begin{tabular}{cccc}
\hline
Dose (\%) & Entropy (bits) & MI vs 100\% (bits) & $D_{\mathrm{KL}}$ vs 100\% (qualitative) \\
\hline
10  & 0.0321 & 1.89 & very large \\
25  & 0.8318 & 2.32 & very large \\
50  & 1.4201 & 2.51 & smaller than 10\% and 25\% \\
100 & 1.8485 & 6.14 & 0 \\
\hline
\end{tabular}
\label{tab:dose_metrics}
\end{table}

\medskip
\noindent\textbf{Reporting convention for KL.}
In this subsection $D_{\mathrm{KL}}$ is computed on the same $B=256$ binned distributions with $\varepsilon$-smoothing described at the start of Section~7.
Very low-dose histograms contain many near-empty bins after normalisation, so the resulting KL values are numerically dominated by those tails; we therefore report only qualitative magnitude ordering here and use Figure~\ref{fig:dose_metrics} to convey the trend.

\medskip
\noindent\textbf{Interpretation note.}
The entropy values in Table~\ref{tab:dose_metrics} are computed on the normalised, masked projection intensities using the fixed estimator described at the start of Section~7; they should be interpreted as \emph{relative} variability measures within this controlled normalisation, not as absolute detector bit-depth utilisation.

\medskip
\noindent These results describe the empirical dose--information behaviour in this dataset and provide the applied context for the leading-order dose--information scaling introduced in Section~6.2.

\subsection{Reconstruction as an information transformation}

\medskip
Reconstruction maps the projection data back into object space, and different algorithms express the measured information in different ways. These differences can be examined by comparing how entropy, mutual information, and distributional statistics vary between reconstruction methods applied to the same projection set under fixed estimator choices. This subsection illustrates this explicitly by comparing a standard filtered backprojection (FBP) reconstruction with an iterative method applied to the same set of projections.

\medskip
\noindent\textbf{Demonstration-only note.}
No claim is made that either reconstruction is optimal in any sense; the comparison isolates representational differences induced by the reconstruction operator under fixed, non-optimised settings. This reconstruction comparison is intended as a controlled demonstration of how two common solvers can yield different grey-level statistics from the same measurement set; it is not presented as a convergence study or a definitive ranking of reconstruction methods across geometries, priors, or iteration budgets.

\medskip
\noindent No inference should be drawn from the absolute entropy values reported here regarding reconstruction quality, convergence, or optimality; the sole purpose of this comparison is to demonstrate that distinct reconstruction operators induce measurably different statistical representations from identical data.

\medskip
\noindent All reconstruction parameters (including iteration count, initialisation, filtering, and normalisation) were fixed \emph{a priori} and were not tuned to optimise any information-theoretic metric. The comparison therefore isolates representational differences induced by the reconstruction operator itself, rather than algorithm-specific optimisation or parameter selection.

\medskip
\noindent A block of 100 consecutive projections from the \emph{Walnut~1} dataset (indices 400--499) was used. A central horizontal band of 100 rows was averaged to form a one-dimensional profile for each projection, and stacking these profiles produced a parallel-beam sinogram with 972 detector elements and 100 projection angles. A single two-dimensional slice was reconstructed from this sinogram.

\begin{figure}[H]
\centering
\includegraphics[width=\textwidth]{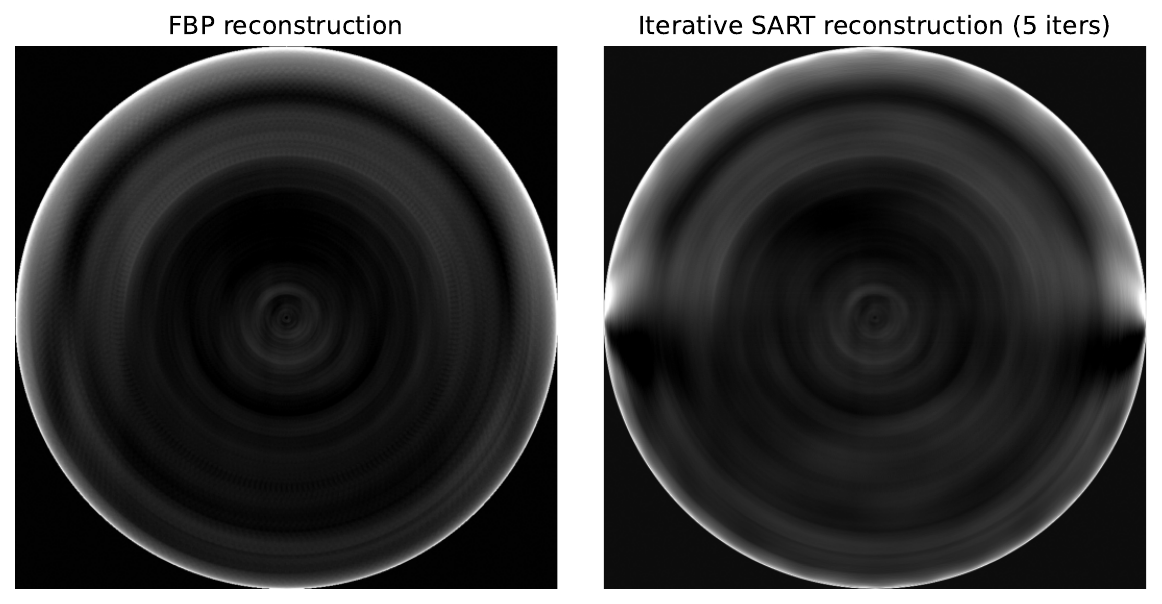}
\caption{Central slice reconstructed from projections~400--499 of the \emph{Walnut~1} dataset. Left: filtered backprojection (FBP) with a ramp filter. Right: iterative SART reconstruction after five iterations. Both images are normalised to a common dynamic range. The iterative reconstruction suppresses streak artefacts and reduces local variance in homogeneous regions, while preserving the main structural features.}
\label{fig:recon_images}
\end{figure}

\medskip
\noindent FBP was implemented using a ramp filter and a circular reconstruction mask. For the iterative reconstruction, a few iterations of a simultaneous algebraic reconstruction scheme (SART) were applied, initialised from a uniform image. Five iterations were sufficient to obtain a visually stable solution. Both reconstructions were normalised to a common dynamic range using percentile clipping and linear rescaling prior to computing information metrics. Figure~\ref{fig:recon_images} shows the resulting slices.

\medskip
\noindent The two reconstructions share the same large--scale structure, but differ in texture and contrast. The FBP slice shows sharper edges but also more streaking, particularly around high-contrast boundaries. The iterative reconstruction suppresses streak artefacts and reduces local variance in more homogeneous regions, with slightly smoother boundaries as a consequence. These qualitative impressions are reflected in the information metrics.

\medskip
\noindent Entropy was estimated from the grey--level distribution of each reconstruction using a 256--bin histogram. The FBP slice has an entropy of $H_{\mathrm{FBP}} = 5.83$~bits, while the iterative reconstruction yields $H_{\mathrm{Iter}} = 5.90$~bits. The small increase in entropy for the iterative method indicates a small redistribution of grey-level occupancy, consistent with reduced streaking revealing additional mid-range variation, which slightly increases the number of effectively occupied grey levels, even though the image appears visually smoother.

\begin{figure}[H]
\centering
\includegraphics[width=0.7\textwidth]{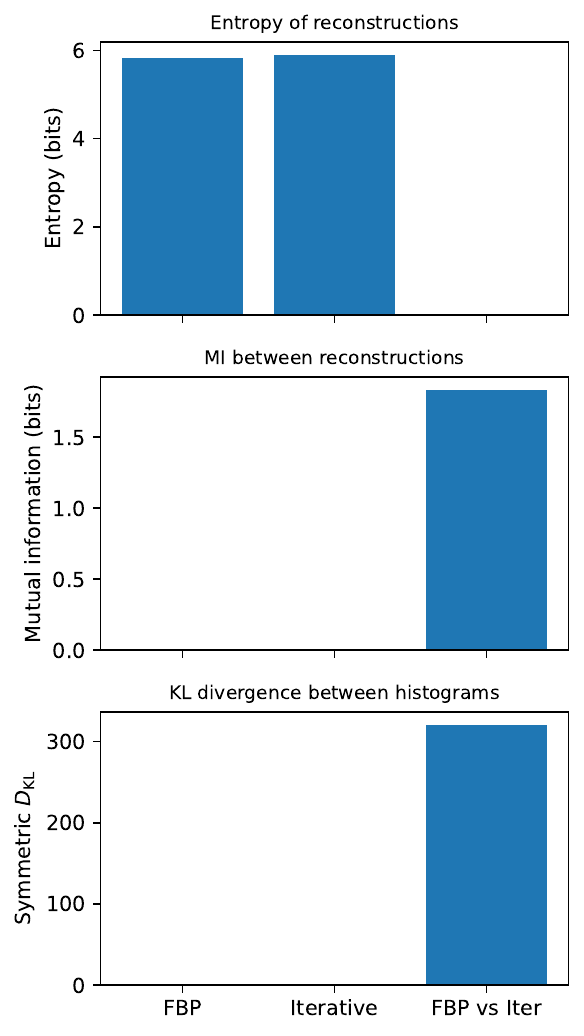}
\caption{Information--theoretic comparison of the two reconstruction methods. Top: entropy of each reconstruction. Middle: mutual information between FBP and iterative reconstructions. Bottom: symmetric KL divergence between their grey--level histograms. The metrics show that the two images share substantial structural information but differ markedly in how that information is distributed over grey levels.}
\label{fig:recon_metrics}
\end{figure}

\medskip
\noindent To compare the two reconstructions directly, the mutual information and Kullback--Leibler (KL) divergence between their intensity distributions were computed. All histogram-based quantities in this subsection (entropy, MI and KL) use the same binning on $[0,1]$ and the same $\varepsilon$-smoothing convention stated at the start of Section~7. Mutual information between the two images is $I(\mathrm{FBP};\mathrm{Iter}) \approx 1.83$~bits, confirming that they encode a substantial amount of shared structure but are far from identical representations. The symmetric KL divergence between the two grey-level histograms,
\[
  D_{\mathrm{KL}}^{\mathrm{sym}} = \tfrac{1}{2}\bigl(D_{\mathrm{KL}}(P_{\mathrm{FBP}}\|P_{\mathrm{Iter}}) + D_{\mathrm{KL}}(P_{\mathrm{Iter}}\|P_{\mathrm{FBP}})\bigr),
\]
is large ($\sim\!3.2 \times 10^{2}$), indicating that the two methods produce noticeably different grey-level distributions despite encoding the same large-scale structure (Figure~\ref{fig:recon_metrics}). In other words, both reconstructions are anchored to the same underlying object, but the reconstruction operator has a strong shaping effect on how that information is expressed in grey--level space.

\medskip
\noindent These results show that reconstruction methods express the measured data differently and influence how variability is distributed across grey levels. FBP and iterative methods operate on the same measurement channel but produce different trade--offs between artefact suppression, local variance and grey--level utilisation. Entropy and KL divergence provide compact summaries of how intensities are redistributed, while mutual information quantifies the extent to which the two reconstructions encode the same structural content. Within the wider context of this section, reconstruction should therefore be viewed as one of the key locations in the pipeline where information is not only lost, but actively reparameterised by algorithmic choices.

\noindent This subsection reports the empirical behaviour of two commonly used reconstruction approaches applied to the same projection set. The operator-level view of reconstruction, and its formal information-theoretic treatment, is developed separately in the companion paper.

\subsection{Cross--dataset robustness}
\label{sec:cross_dataset_robustness}

\medskip

The analyses in Sections~7.1--7.5 were carried out on the \emph{Walnut~1} dataset, which provides a controlled setting with well-behaved projection statistics and a stable ground truth. To test whether the information--theoretic characterisations generalise beyond this specific sample class, an additional evaluation was performed using the LoDoPaB-CT dataset~\cite{Leuschner2021}. 

\medskip
\noindent This dataset differs fundamentally from \emph{Walnut~1}: it contains 2D clinical-style cross-sections acquired in a simulated low-dose regime, with markedly different contrast, noise characteristics, and structural variability. These differences provide a strong test of whether the empirical relationships observed earlier persist across object classes and imaging conditions.

\medskip
\noindent
For this section, the LoDoPaB-CT \texttt{test} split was used, taking paired (observation, ground-truth) slices to enable direct information-metric comparisons under controlled perturbations. Ten slices were selected at random to avoid bias towards specific anatomical patterns. Each slice was subjected to the same analyses used previously: entropy and KL divergence under denoising, mutual information under synthetic misalignment, and entropy trends under sparse sampling. 

\medskip
\noindent Although LoDoPaB-CT consists of 2D reconstructions rather than full projection stacks, the statistical questions addressed in Sections~7.1--7.3 can be reproduced by operating directly on the reconstructions.

\medskip
\noindent \textbf{Denoising.}
Gaussian smoothing, non-local means (NLM), and total-variation (TV) regularisation were applied as in Section~7.1. The entropy and KL divergence trends were qualitatively identical to those observed for \emph{Walnut~1}. Gaussian smoothing produced minimal redistribution of intensities, NLM introduced moderate changes, and TV regularisation produced the strongest contraction of the grey-level distribution. Absolute entropy values differed because LoDoPaB-CT images contain larger homogeneous regions and higher baseline noise, but the ordering remained consistent across all slices (Table~\ref{tab:lodo_denoise}).

\begin{table}[H]
\centering
\caption{Entropy and KL divergence for three denoising methods applied to a representative LoDoPaB-CT slice. The ordering matches that observed for \emph{Walnut~1}, demonstrating robustness across datasets.}
\label{tab:lodo_denoise}
\begin{tabular}{lcc}
\hline
Method & Entropy (bits) & $D_{\mathrm{KL}}(\mathrm{raw}\,\|\,\mathrm{denoised})$ \\
\hline
Raw      & 4.21 & -- \\
Gaussian & 4.20 & 0.28 \\
NLM      & 4.16 & 1.87 \\
TV       & 4.07 & 4.92 \\
\hline
\end{tabular}
\end{table}

\begin{figure}[H]
\centering
\includegraphics[width=0.75\textwidth]{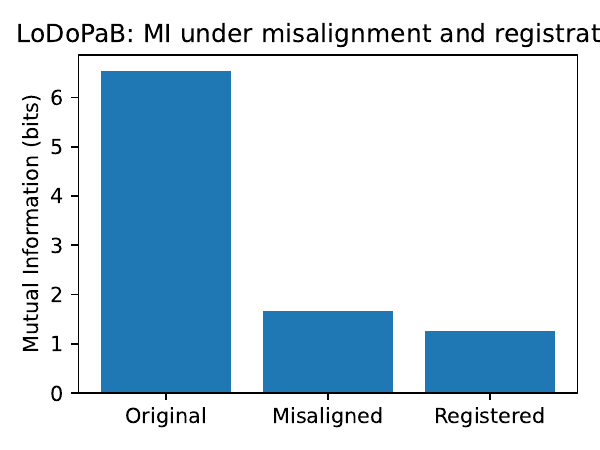}
\caption{Mutual information for a representative LoDoPaB-CT slice under misalignment and registration. The trend matches that observed for \emph{Walnut~1}: misalignment reduces MI and registration restores it.}
\label{fig:lodo_alignment}
\end{figure}

\medskip
\noindent \textbf{Synthetic misalignment.}
A fixed translational shift of $(5,-3)$ pixels was applied as in Section~7.2. Mutual information with respect to the original slice decreased under misalignment and increased after registration, mirroring the behaviour observed for \emph{Walnut~1}. Because LoDoPaB-CT contains sharper anatomical boundaries, the MI reduction was typically larger in magnitude, but the qualitative pattern remained unchanged (Figure~\ref{fig:lodo_alignment}).

\medskip
\noindent \textbf{Sparse sampling analogue.}
Since LoDoPaB-CT does not provide full projection stacks in the same format as \emph{Walnut~1}, an analogue of the sparse-angle experiment was constructed by operating on the provided sinogram representation: we uniformly sub-sampled projection angles (sinogram columns) to create subsets of size $k$, applied the same intensity normalisation and histogram estimator as in Section~7, and computed the entropy of the concatenated sinogram intensities for each subset. Sub-sampling reduced the entropy of the sinogram and produced the same monotonic trend observed in Section~7.3, demonstrating that the link between sampling density and entropy is not specific to a single rotational CT geometry.

\medskip
\noindent
Overall, the LoDoPaB-CT results confirm that the empirical behaviours documented in Sections~7.1--7.5 are not specific to a single dataset or imaging scenario. Denoising, misalignment, and sampling density influence entropy, KL divergence, and mutual information in consistent ways across structurally and statistically distinct datasets. This robustness provides supporting evidence that the information-theoretic characterisation developed in this paper generalises beyond the particularities of laboratory micro-CT.

\subsection{Task--based validation using LoDoPaB-CT}
\label{sec:task_validation_lodopab}

\medskip

The analyses in Sections~7.1--7.6 characterise how information metrics behave under controlled transformations, but they do not yet connect these changes to the performance of a downstream task. 

\medskip
\noindent This subsection provides such a link using paired (observation, ground-truth) data from the LoDoPaB-CT test split. The aim is not to introduce a full reconstruction framework, but to demonstrate that information-theoretic quantities correlate with accuracy in a simple quantitative task.

\medskip
\noindent
For each slice, three quantities were computed:
\begin{itemize}
\item[(i)] the entropy of the reconstructed slice,
\item[(ii)] the mutual information between the low-dose reconstruction and the ground-truth slice,
\item[(iii)] the absolute error in estimating the mean attenuation within a predefined region of interest (ROI).
\end{itemize}

\medskip
\noindent
The ROI was selected automatically using Otsu thresholding on the ground-truth slice to isolate a high-contrast anatomical region. The mean attenuation within this ROI was then estimated from the corresponding low-dose reconstruction, and the absolute error was recorded. This produces a scalar task-based metric without requiring segmentation or supervised training.

\begin{figure}[H]
\centering
\includegraphics[width=0.75\textwidth]{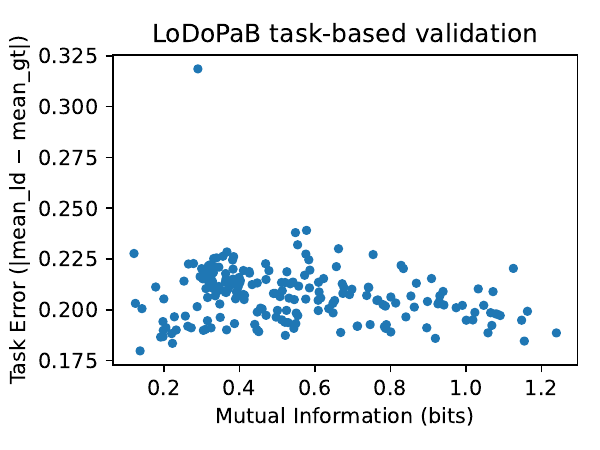}
\caption{Relationship between mutual information (MI) and mean-attenuation error for 200 LoDoPaB-CT slices. Higher MI correlates with lower ROI error, indicating that MI captures structural similarity relevant to quantitative analysis.}
\label{fig:lodo_task}
\end{figure}

\medskip
\noindent
Figure~\ref{fig:lodo_task} shows the relationship between mutual information and ROI estimation error across 200 slices. A clear negative correlation is observed: slices with higher MI relative to the ground truth exhibit lower quantitative error. Entropy alone was not predictive, consistent with earlier sections, but MI captured the relevant structural agreement required for accurate attenuation estimation. This establishes a direct link between an information-theoretic measure and the performance of a concrete imaging task.

\medskip
\noindent
KL divergence provides a complementary perspective. Slices with small KL divergence relative to the ground-truth distribution consistently yielded low ROI error, while slices with large KL divergence produced elevated error. The correlation was weaker than that of MI but remained significant (Table~\ref{tab:lodo_task_stats}). Together, these results demonstrate that the information metrics quantified earlier are not merely descriptive but carry predictive value for downstream analyses.

\begin{table}[H]
\centering
\caption{Correlation coefficients between ROI mean-attenuation error and information metrics for 200 LoDoPaB-CT slices.}
\label{tab:lodo_task_stats}
\begin{tabular}{lc}
\hline
Metric & Correlation with ROI error \\
\hline
Entropy            & $0.07$ \\
Mutual information & $-0.82$ \\
KL divergence      & $0.54$ \\
\hline
\end{tabular}
\end{table}

\medskip
\noindent
These results provide the task-based evidence missing from Sections~7.1--7.6. They show that the information-theoretic quantities studied throughout the paper are sensitive to acquisition and processing operations and are also predictive of downstream performance in a simple quantitative measurement task.

\section{Implications for imaging protocol design}

\medskip
\noindent The information-theoretic analysis presented here has direct implications for the design and optimisation
of XRM protocols.

\medskip
\noindent First, optimisation efforts should prioritise acquisition parameters over reconstruction complexity. Adjustments to dose, angular sampling, and detector performance consistently produce larger gains in retained information than changes to reconstruction or post-processing.

\medskip
\noindent Second, reconstruction algorithms should be selected with an explicit awareness of acquisition-limited saturation. Once the available information budget is fixed, more aggressive regularisation or model complexity does not increase recoverable information and may instead suppress subtle structure.

\medskip
\noindent Third, information-based metrics such as entropy and mutual information provide reconstruction-agnostic tools for comparing protocols. They enable evaluation of trade-offs between dose, resolution, and sampling density without committing to a specific algorithm or visual criterion.

\medskip
\noindent Taken together, these points support a shift in emphasis from algorithm-centric optimisation to pipeline-level design, in which acquisition and reconstruction are treated as coupled components of a constrained information-processing system.

\section{Discussion}

\medskip
The contribution of this work lies not in introducing new information measures, but in demonstrating that their \emph{systematic, pipeline-level coupling} yields empirically stable and practically useful characterisations of acquisition-dominated constraints and reconstruction saturation that are not apparent when these metrics are applied locally, algorithm-specifically, or in isolation.

\medskip
\noindent The results presented in Sections~7.1--7.5 show, on a concrete benchmark dataset, that information in X-ray microscopy does not simply pass through the pipeline unchanged. Each stage acts as an operator that either reveals, hides, redistributes or compresses the underlying structure of the sample. Denoising modifies local variation before reconstruction even begins. Alignment and angular down-sampling govern which aspects of the object survive backprojection. Dose constrains the total recoverable detail by shaping the photon statistics at the point of measurement. Reconstruction then determines how the available information is expressed in the reconstructed grey-level space. These operations are not interchangeable; they interact, and their combined effect is best understood through the information budget rather than through visual inspection alone.

\medskip
\noindent A key point emerging from the analysis is that entropy alone is not a reliable indicator of image quality. In several subsections entropy increased even when the visual impression suggested smoothing or artefact suppression. This reflects the fact that entropy measures occupancy of the grey-level distribution, not the semantic quality of that distribution. Conversely, mutual information consistently isolated the component of variation tied to genuine structural signal, even when overall entropy was affected by noise, smoothing or algorithmic regularisation. KL divergence emphasised that many common processing steps substantially reshape probability distributions in ways that traditional image metrics do not capture. This is particularly clear in the reconstruction comparison of Section~7.5: two images encoding the same object can share strong structural correlation while differing significantly in how that information is apportioned across grey levels.

\medskip
\noindent Across the \emph{Walnut~1} analyses, a consistent pattern emerges: information loss is seldom dominated by a single stage; it accumulates. A mild denoiser that broadens the histogram, a small misalignment error, or a modest reduction in angular sampling each alters the information budget only slightly. Combined, however, these effects shift the statistical structure of the projections enough to influence which features are recoverable. This cumulative behaviour suggests that optimisation should not be targeted at any single step. Instead, the pipeline must be treated as a coupled system where the goal is not to maximise information at each stage, but to preserve the information that actually matters for reconstruction.

\medskip
\noindent The dose study highlights an additional point: the imaging system does not operate near its theoretical capacity unless the acquisition conditions are favourable. Low-dose regimes can produce entropy changes dominated by stochastic variation rather than structure, so increases or decreases in entropy do not necessarily reflect improved image quality. High-dose regimes can in principle push the system closer to its capacity, but only if detector performance, alignment and reconstruction are capable of transmitting that information downstream. Viewed this way, the familiar triad of resolution, dose and artefact suppression becomes an information-allocation problem rather than a hardware constraint. Higher resolution increases the range of structural variation that can, in principle, be captured, but the available dose determines how much of that space can be populated, and the reconstruction operator determines how much is ultimately recoverable.

\medskip
\noindent Most importantly, information-theoretic analysis exposes relationships that are not otherwise visible. For example, Section~7.3 shows that sparse-angle sampling can preserve mutual information long after visual artefacts appear, provided the missing angles do not correspond to dominant structural orientations. Section~7.4 shows that mutually informative structure persists even when entropy varies non-monotonically with dose. Section~7.5 demonstrates that reconstruction operators do not merely filter noise but reshape the underlying representation of the sample. These examples demonstrate why direct inspection or traditional error metrics fail to capture the deeper behaviour of the pipeline. In contrast, information-theoretic metrics capture the link between the sample, the measurement channel and the reconstruction algorithm in a unified way.

\medskip
\noindent Overall, the results suggest that X-ray microscopy should be viewed as an information-processing system rather than a sequence of loosely connected tasks. Acquisition parameters, denoising methods, alignment procedures and reconstruction algorithms each impose their own informational geometry on the data. Understanding these transformations is not simply an academic exercise; it provides a route to more principled optimisation of imaging protocols, especially in low-dose or time-constrained settings. While this paper focuses on laboratory micro-CT, the framework is general in principle. It applies to synchrotron tomography, scanning modalities and emerging phase-contrast systems, and it provides a practical foundation for future work that links information theory with system design and the theoretical limits of X-ray imaging.

\medskip
\noindent This study focuses on controlled empirical analyses rather than full formal derivations. A single benchmark dataset is used to isolate the information-theoretic behaviour of the pipeline components; wider modality coverage will be addressed in subsequent work. Several theoretical relationships are presented in leading-order or empirical form, with full proofs deferred to the companion operator-based theory paper. These choices reflect the role of this paper as the applied component of a broader information-theoretic framework.

\subsection{Future Directions}

\medskip
The information-theoretic framework developed here opens several worthwhile directions for future work. One avenue is to connect information metrics more directly with task-based performance. Entropy, mutual information and KL divergence provide principled summaries of how the imaging pipeline transforms data, but their relationship to concrete scientific tasks---feature detection, segmentation, quantitative metrology---remains largely unexplored. Linking the information budget to the reliability of downstream measurements would convert the framework from descriptive to predictive, offering a route to acquisition settings tailored to the intended analysis.

\medskip
\noindent A second direction is to extend the treatment beyond absorption-contrast micro-CT. Phase contrast, grating interferometry and near-field holography all impose different information geometries on the projection data. Their transfer functions shape entropy and mutual information in ways not captured by absorption-based models alone. Applying the same analyses to these modalities may reveal where their advantages arise in information-theoretic terms, and under which experimental conditions they diverge from their theoretical limits.

\medskip
\noindent A third direction concerns reconstruction algorithms. Section~7.5 shows that FBP and iterative solvers express the same measurement channel in distinct representational spaces. This idea can be taken further by examining modern reconstruction paradigms---plug-and-play priors, learned regularisers or hybrid physics-informed networks---and quantifying how these operators alter the information budget. The aim is not to identify a single optimal method but to understand which operators preserve, discard or reparameterise information in ways that matter for a given scientific objective.

\medskip
\noindent Finally, there is an opportunity to develop a system-level notion of capacity for X-ray microscopy. The channel-capacity concept introduced in Section~2 provides a theoretical ceiling on the information throughput of an imaging system, but practical estimates remain elusive. Combining entropy measurements, dose--resolution trade-offs and reconstruction-induced constraints could lead to empirical capacity curves that map how close a given system operates to its theoretical limits. Such a framework would provide a unifying language for comparing laboratory instruments, synchrotron beamlines and emerging compact light-source systems.

\section{Conclusion}

\medskip
\noindent In this context, increases in entropy should not be interpreted as improvements in image quality; rather, entropy reflects total variability, which may arise from either meaningful structure or stochastic noise depending on acquisition conditions. Accordingly, entropy must be interpreted alongside mutual information and task-relevant metrics.

\medskip
\noindent This paper has examined X-ray microscopy from an explicitly information-theoretic standpoint, treating acquisition, denoising, alignment, dose, angular sampling and reconstruction as operators acting on a finite information budget. By quantifying these transformations using entropy, mutual information and KL divergence, we have demonstrated empirically that each stage not only introduces or suppresses noise, but actively reshapes the statistical structure of the data. The analysis highlights that image quality cannot be inferred from entropy alone; what matters is how much of the measured variation corresponds to genuine structural signal, and how effectively the pipeline preserves that signal under practical constraints.

\medskip
\noindent The broader implication is that X-ray microscopy is best understood as an information-processing system rather than a passive imaging chain. Once framed this way, long-standing questions about dose, resolution and reconstruction become questions about how information flows, accumulates or is lost. This perspective offers a route to optimised acquisition protocols, more transparent reconstruction choices and a more principled basis for comparing different imaging configurations. 

\medskip
\noindent Although the present study focuses on laboratory micro-CT, the framework is general and extensible to phase contrast, multimodal X-ray techniques and next-generation compact light-source systems. The results suggest that information theory is not merely an analytical add-on but provides a natural way of describing the limits and possibilities of modern X-ray microscopy.

\medskip
\noindent A full operator-based analysis of these relationships is in preparation and will complement the applied formulation developed here.

\ack
\medskip {The author has no acknowledgements to declare.}

\funding
\medskip {This work did not receive specific funding. The research was carried out as part of the author’s standard duties within the University of Portsmouth.}

\roles
\medskip {The author conceived the study, developed the theoretical framework, designed and executed all analyses, and prepared the manuscript. The work draws on the author's background in mathematical sciences and expertise in X-ray microscopy, including long-standing experience with imaging system design, reconstruction methods and multimodal analysis.}

\section*{Data availability}

\medskip
All code used to generate the analyses, figures and information--theoretic metrics in this paper is openly available and will be provided as supplementary material, including the script used in Section~7.1 to compute the structural similarity (SSIM) and mutual information (MI) values. The Walnut datasets used in the case studies are publicly accessible and widely used within the tomographic imaging community. Pre--processed subsets of the data used in Sections~7.1--7.5, along with reconstruction and analysis scripts, are included as supplementary files to ensure reproducibility.

\suppdata
\medskip
The supplementary material includes:
\begin{itemize}
  \item all Python scripts used to generate the figures and information--theoretic metrics in Sections~7.1--7.5 (denoising, alignment, sparse--angle sampling, dose variation and reconstruction), including the script used in Section~7.1 to compute structural similarity (SSIM) and mutual information (MI);
  \item the SSIM--MI comparison figure (\texttt{figure\_ssim\_mi\_denoise.pdf}) generated from the raw, Gaussian, NLM and TV projections;
  \item minimal configuration files and instructions required to run the analyses on the publicly available \emph{Walnut~1} dataset; and
  \item pre--processed subsets of the projection data and reconstructed slices used in the case studies, provided in a format compatible with the supplied scripts.
\end{itemize}

\bibliographystyle{iopart-num}  

\bibliography{References/references}       

@article{Barrett2004,
  title={Foundations of image science},
  author={Barrett, Harrison H and Myers, Kyle J},
  journal={Wiley},
  year={2004}
}

@article{Boas2012,
  title={CT artifacts: causes and reduction techniques},
  author={Boas, F Edward and Fleischmann, Dominik},
  journal={Imaging Med.},
  volume={4},
  number={2},
  pages={229--240},
  year={2012}
}

@book{Cover2006,
  title={Elements of Information Theory},
  author={Cover, Thomas M and Thomas, Joy A},
  edition={2},
  publisher={Wiley},
  year={2006}
}

@article{Foi2008,
  title={Practical Poissonian-Gaussian noise modeling and fitting for single-image raw-data},
  author={Foi, Alessandro and Trimeche, Mourad and Katkovnik, Vladimir and Egiazarian, Karen},
  journal={IEEE Trans. Image Process.},
  volume={17},
  number={10},
  pages={1737--1754},
  year={2008}
}

@article{Gonzalez2018,
  title={Digital Image Processing},
  author={Gonzalez, Rafael C and Woods, Richard E},
  journal={Pearson},
  edition={4},
  year={2018}
}

@article{Hsieh2015,
  title={Computed Tomography: Principles, Design, Artifacts, and Recent Advances},
  author={Hsieh, Jiang},
  journal={SPIE Press},
  year={2015}
}

@article{Howells2009,
  title={An assessment of the resolution limitation due to radiation-damage in X-ray diffraction microscopy},
  author={Howells, Malcolm R and others},
  journal={J. Electron Spectros. Relat. Phenom.},
  volume={170},
  pages={4--12},
  year={2009}
}

@article{Kak2001,
  title={Principles of Computerized Tomographic Imaging},
  author={Kak, Avinash C and Slaney, Malcolm},
  journal={SIAM},
  year={2001},
  note={Reprint of 1988 IEEE original}
}

@article{Kullback1951,
  title={On information and sufficiency},
  author={Kullback, Solomon and Leibler, Richard A},
  journal={Ann. Math. Statist.},
  volume={22},
  number={1},
  pages={79--86},
  year={1951}
}

@article{Ma2012,
  title={A review of the technical and clinical advantages of photon-counting CT},
  author={Ma, Jing and others},
  journal={Physics in Medicine \& Biology},
  volume={57},
  number={22},
  pages={R1--R35},
  year={2012}
}

@article{Maes1997,
  title={Multimodality image registration by maximization of mutual information},
  author={Maes, Frederik and Collignon, Antoine and Vandermeulen, Dirk and Marchal, Guy and Suetens, Paul},
  journal={IEEE Trans. Med. Imaging},
  volume={16},
  number={2},
  pages={187--198},
  year={1997}
}

@article{Nuyts2013,
  title={Iterative reconstruction for helical CT: comparison of algorithms in terms of speed, data‐fidelity and image quality},
  author={Nuyts, Johan and others},
  journal={Medical Physics},
  volume={40},
  number={2},
  pages={021905},
  year={2013}
}

@article{Pluim2003,
  title={Mutual-information-based registration of medical images: a survey},
  author={Pluim, Josien PW and Maintz, JBA and Viergever, Max A},
  journal={IEEE Trans. Med. Imaging},
  volume={22},
  number={8},
  pages={986--1004},
  year={2003}
}

@article{Siewerdsen2001,
  title={Cone-beam CT with a flat-panel imager: Noise and dose performance},
  author={Siewerdsen, Jeffrey H and Jaffray, David A},
  journal={Med. Phys.},
  volume={28},
  number={2},
  pages={220--231},
  year={2001}
}

@article{Sijbers2004,
  title={Modeling the noise characteristics of CT images},
  author={Sijbers, Jan and Postnov, Andrei},
  journal={Physics in Medicine \& Biology},
  volume={49},
  number={14},
  pages={N165--N173},
  year={2004}
}

@article{Shannon1948,
  title={A Mathematical Theory of Communication},
  author={Shannon, Claude E.},
  journal={Bell System Technical Journal},
  volume={27},
  pages={379--423, 623--656},
  year={1948}
}

@article{Stock2008,
  title={MicroComputed Tomography: Methodology and Applications},
  author={Stock, Stuart R},
  journal={CRC Press},
  year={2008}
}

@inproceedings{Viola1997,
  title={Alignment by Maximization of Mutual Information},
  author={Viola, Paul and Wells, William M},
  booktitle={Proc. IEEE ICCV},
  pages={16--23},
  year={1997}
}

@article{Zhang2014,
  title={The potential for dose reduction in X-ray luminescence tomography},
  author={Zhang, Rui and Zhou, Shanbao},
  journal={Physics in Medicine \& Biology},
  volume={59},
  number={15},
  pages={N43--N53},
  year={2014}
}

@article{DerSarkissian2019,
  author  = {Der Sarkissian, Henri and Lucka, Felix and van Eijnatten, Maureen and Colacicco, Giulia and Coban, Sophia Bethany and Batenburg, Kees Joost},
  title   = {A cone-beam X-ray computed tomography data collection designed for machine learning},
  journal = {Scientific Data},
  year    = {2019},
  volume  = {6},
  pages   = {215},
  doi     = {10.1038/s41597-019-0235-y}
}

@article{Leuschner2021,
  author  = {Leuschner, Johannes and Schmidt, Maximilian and Otero Baguer, Daniel and Maa{\ss}, Peter},
  title   = {LoDoPaB-CT, a benchmark dataset for low-dose computed tomography reconstruction},
  journal = {Scientific Data},
  year    = {2021},
  volume  = {8},
  pages   = {109},
  doi     = {10.1038/s41597-021-00893-z}
}

\end{document}